\documentclass[journal=ancac3,manuscript=article]{achemso}

\usepackage[version=3]{mhchem} 
\usepackage{color}
\usepackage{wrapfig}
\usepackage[labelfont=bf]{caption}

\newlength{\mylen}
\setbox1=\hbox{$\bullet$}\setbox2=\hbox{\tiny$\bullet$}
\usepackage{textcomp}
\usepackage{lastpage}
\usepackage{fancyhdr}
\usepackage[T1]{fontenc}
\usepackage[utf8]{inputenc}

\usepackage{amsmath}

\usepackage{color}
\definecolor{darkgreen}{rgb}{0.2,0.2,0.7} 
\definecolor{darkblue}{rgb}{0.2,0.2,0.7} 
\definecolor{darkred}{rgb}{0.2,0.2,0.7} 

\usepackage{ragged2e}
\usepackage[bookmarks=true,pdfborder={0 0 0}]{hyperref}
\hypersetup{
  linkcolor  = darkblue,
  citecolor  = darkblue, 
  urlcolor   = darkblue, 
  colorlinks = true,
}
\usepackage{units}
\SectionNumbersOn

\usepackage{achemso}
\setkeys{acs}{maxauthors = 0}
\author{Clarice D. Aiello}
\email{cla@ucla.edu}
\affiliation{California NanoSystems Institute, University of California, Los Angeles, Los Angeles, CA, USA, 90095}
\alsoaffiliation{Department of Electrical and Computer Engineering, University of California, Los Angeles, Los Angeles, CA, USA, 90095}

\author{Muneer Abbas}
\affiliation{Department of Microbiology, Howard University, Washington, DC, USA, 20059}

\author{John M.\ Abendroth}
\affiliation{Laboratory for Solid State Physics, ETH Zürich, Zürich 8093}

\author{Andrei Afanasev}
\affiliation{George Washington University, Washington, DC, 20052, USA}

\author{Shivang Agarwal}
\affiliation{Department of Electrical and Computer Engineering, University of California, Los Angeles, Los Angeles, CA, USA, 90095}

\author{Amartya S. Banerjee}
\affiliation{Department of Materials Science and Engineering, University of California, Los Angeles, Los Angeles, CA, USA, 90095}
\alsoaffiliation{California NanoSystems Institute, University of California, Los Angeles, Los Angeles, CA, USA, 90095}

\author{David N. Beratan}
\affiliation{Departments of Chemistry, Biochemistry, and Physics, Duke University, Durham, NC, USA, 27708}

\author{Jason N. Belling}
\affiliation{California NanoSystems Institute, University of California, Los Angeles, Los Angeles, CA, USA, 90095}
\alsoaffiliation{Department of Chemistry & Biochemistry, University of California, Los Angeles, Los Angeles, CA, 
USA, 90095}

\author{Bertrand Berche}
\affiliation{Laboratoire de Physique et Chimie Th\'e oriques, UMR Universit\'e de Lorraine-CNRS, 7019 54506 Vand\oeuvre les Nancy, France}

\author{Antia Botana}
\affiliation{Department of Physics, Arizona State University, Tempe, AZ, USA, 85287}

\author{Justin R. Caram}
\affiliation{Department of Chemistry & Biochemistry, University of California, Los Angeles, Los Angeles, CA, USA, 90095}

\author{Giuseppe Luca Celardo}
\affiliation{Institute of Physics, Benemerita Universidad Autonoma de Puebla, Apartado Postal J-48,  72570, Mexico}

\author{Gianaurelio Cuniberti}
\affiliation{Institute for Materials Science and Max Bergmann Center of Biomaterials, Dresden University of Technology, 01062 Dresden, Germany}

\author{Aitzol Garcia-Etxarri}
\affiliation{Donostia International Physics Center,	Euskal Herriko Unibertsitatea,	20018 Donostia, Euskadi, Spain}

\author{Arezoo Dianat}
\affiliation{Institute for Materials Science and Max Bergmann Center of Biomaterials, Dresden University of Technology, 01062 Dresden, Germany}

\author{Ismael Diez-Perez}
\affiliation{Department of Chemistry, Faculty of Natural and Mathematical Sciences
King's College London, 7 Trinity Street, London, SE1 1DB, UK}

\author{Yuqi Guo}
\affiliation{School for Engineering of Matter, Transport and Energy, Arizona State University, Tempe, AZ, USA, 85287}

\author{Rafael Gutierrez}
\affiliation{Institute for Materials Science and Max Bergmann Center of Biomaterials, Dresden University of Technology, 01062 Dresden, Germany}

\author{Carmen Herrmann}
\affiliation{Department of Chemistry, University of Hamburg, 20146 Hamburg, Germany}

\author{Joshua Hihath}
\affiliation{Department of Electrical and Computer Engineering, University of California, Davis, Davis, CA, USA, 95616}

\author{Suneet Kale}
\affiliation{School of Molecular Sciences, Arizona State University, Tempe, AZ, USA, 85287}

\author{Philip Kurian}
\affiliation{Quantum Biology Laboratory, Graduate School, Howard University, Washington, DC, USA 20059}

\author{Ying-Cheng Lai}
\affiliation{School of Electrical, Computer and Energy Engineering, Arizona State University, Tempe, AZ, USA, 85287}

\author{Alexander Lopez}
\affiliation{Escuela Superior Polit\'ecnica del Litoral, ESPOL, Campus Gustavo Galindo Km. 30.5 V\'ia Perimetral, PO Box 09-01-5863, Guayaquil, Ecuador}

\author{Ernesto Medina}
\affiliation{School of Physical Sciences & Nanotechnology, Yachay Tech University, 100119-Urcuquí, Ecuador}

\author{Vladimiro Mujica}
\affiliation{School of Molecular Sciences, Arizona State University, Tempe, AZ, USA, 85287}

\author{Ron Naaman}
\affiliation{Department of Chemical and Biological Physics, Weizmann Institute of Science, Rehovot 76100, Israel}

\author{Mohammadreza Noormandipour}
\affiliation{Department of Electrical and Computer Engineering, University of California, Los Angeles, Los Angeles, CA, USA, 90095}
\alsoaffiliation{TCM Group, Cavendish Laboratory, University of Cambridge, J.J.~Thomson Avenue, Cambridge CB3 0HE, United Kingdom}

\author{Julio L. Palma}
\affiliation{Department of Chemistry, Pennsylvania State University, Lemont Furnace, PA, USA, 15456}

\author{Yossi Paltiel}
\affiliation{Applied Physics Department and the Center for Nano-Science and Nano-Technology, Hebrew University of Jerusalem, Jerusalem 91904, Israel}

\author{William T. Petuskey}
\affiliation{School of Molecular Sciences, Arizona State University, Tempe, AZ, USA, 85287}

\author{Jo\~ao Carlos Ribeiro-Silva}
\affiliation{Laboratory of Genetics and Molecular Cardiology, Heart Institute, University of S\~ao Paulo Medical School, S\~ao Paulo, Brazil}

\author{Juan José Saenz}
\affiliation{Donostia International Physics Center,	Euskal Herriko Unibertsitatea,	20018 Donostia, Euskadi, Spain}

\author{Elton J. G. Santos}
\affiliation{Institute for Condensed Matter Physics and Complex Systems, School of Physics and Astronomy, The University of Edinburgh, EH9 3FD, UK}

\author{Maria Solyanik}
\affiliation{George Washington University, Washington, DC, 20052, USA}

\author{Volker J. Sorger}
\affiliation{George Washington University, Washington, DC, 20052, USA}

\author{Dominik M. Stemer}
\affiliation{California NanoSystems Institute, University of California, Los Angeles, Los Angeles, CA, USA, 90095}
\alsoaffiliation{Department of Materials Science and Engineering, University of California, Los Angeles, Los Angeles, CA, USA, 90095}

\author{Jesus M. Ugalde}
\affiliation{Kimika Fakultatea,	Euskal Herriko Unibertsitatea, 	20080 Donostia, Euskadi, Spain}

\author{Ana Valdes-Curiel}
\affiliation{California NanoSystems Institute, University of California, Los Angeles, Los Angeles, CA, USA, 90095}
\alsoaffiliation{Department of Electrical and Computer Engineering, University of California, Los Angeles, Los Angeles, CA, USA, 90095}

\author{Solmar Varela}
\affiliation{School of Chemical Sciences & Engineering, Yachay Tech University, 100119-Urcuquí, Ecuador}

\author{David H. Waldeck}
\affiliation{Department of Chemistry, University of Pittsburgh, Pittsburgh, PA, USA, 15260}

\author{Paul S. Weiss}
\affiliation{California NanoSystems Institute, University of California, Los Angeles, Los Angeles, CA, USA, 90095}
\alsoaffiliation{Department of Materials Science and Engineering, University of California, Los Angeles, Los Angeles, CA, USA, 90095}

\author{Helmut Zacharias}
\affiliation{Center for Soft Nanoscience, University of M\"unster, 48149 M\"unster, Germany}

\author{Qing Hua Wang}
\affiliation{School for Engineering of Matter, Transport and Energy, Arizona State University, Tempe, AZ, USA, 85287}
\email{qhwang@asu.edu}

\makeatletter
\let\acs@address@list\relax
\makeatother

\title {A Chirality-Based Quantum Leap}


\begin{document}




%
$^\dagger$ California NanoSystems Institute, University of California, Los Angeles, Los Angeles, CA, USA, 90095

$^\ddagger$ Department of Electrical and Computer Engineering, University of California, Los Angeles, Los Angeles, CA, USA, 90095

$^\P$ Department of Microbiology, Howard University, Washington, DC, USA, 20059

$^\S$ Laboratory for Solid State Physics, ETH Zürich, Zürich 8093, Switzerland

$^\parallel$ George Washington University, Washington, DC, 20052, USA

$^\perp$ Department of Materials Science and Engineering, University of California, Los Angeles, Los Angeles, CA, USA, 90095

$^\#$ Department of Chemistry, Duke University, Durham, NC, USA, 27708

$^@$ Department of Chemistry $\&$ Biochemistry, University of California, Los Angeles, Los Angeles, CA, USA, 90095

$^\triangle$ Laboratoire de Physique et Chimie Théoriques, UMR Universit\'e de Lorraine-CNRS, 7019 54506 Vandœuvre les Nancy, France

$^\nabla$ Department of Physics, Arizona State University, Tempe, AZ, USA, 85287

$^{\dagger\dagger}$ Institute of Physics, IFUAP-BUAP, 72000 Puebla, Mexico

$^{\ddagger\ddagger}$  Institute for Materials Science and Max Bergmann Center of Biomaterials, Dresden University of Technology, 01062 Dresden, Germany

$^{\P\P}$ Donostia International Physics Center, Euskal Herriko Unibertsitatea,	20018 Donostia, Euskadi, Spain

$^{\S\S}$ Department of Chemistry, Faculty of Natural and Mathematical Sciences
King's College London, 7 Trinity Street, London, SE1 1DB, UK

$^{\parallel\parallel}$ School for Engineering of Matter, Transport and Energy, Arizona State University, Tempe, AZ, USA, 85287

$^{\perp\perp}$ Department of Chemistry, University of Hamburg, 20146 Hamburg, Germany

$^{\#\#}$Department of Electrical and Computer Engineering, University of California, Davis, Davis, CA, USA, 95616

$^{@@}$ School of Molecular Sciences, Arizona State University, Tempe, AZ, USA, 85287

$^{\triangle\triangle}$ Quantum Biology Laboratory, Graduate School, Howard University, Washington, DC, USA 20059

$^{\nabla\nabla\nabla}$ School of Electrical, Computer and Energy Engineering, Arizona State University, Tempe, AZ, USA, 85287

$^{\dagger\dagger\dagger}$ Escuela Superior Polit\'ecnica del Litoral, ESPOL, Campus Gustavo Galindo Km. 30.5 V\'ia Perimetral, PO Box 09-01-5863, Guayaquil, Ecuador

 $^{\ddagger\ddagger\ddagger}$ School of Physical Sciences $\&$ Nanotechnology, Yachay Tech University, 100119-Urcuquí, Ecuador

$^{\P\P\P}$ Department of Chemical and Biological Physics, Weizmann Institute of Science, Rehovot 76100, Israel

$^{\S\S\S}$ TCM Group, Cavendish Laboratory, University of Cambridge, J.J.~Thomson Avenue, Cambridge CB3 0HE, United Kingdom

$^{\parallel\parallel\parallel}$ Department of Chemistry, Pennsylvania State University, Lemont Furnace, PA, USA, 15456

$^{\perp\perp\perp}$ Applied Physics Department and the Center for Nano-Science and Nano-Technology, Hebrew University of Jerusalem, Jerusalem 91904, Israel

$^\#\#\#$ Laboratory of Genetics and Molecular Cardiology, Heart Institute, University of S\~ao Paulo Medical School, S\~ao Paulo, Brazil

$^{@@@}$ Institute for Condensed Matter Physics and Complex Systems, School of Physics and Astronomy, The University of Edinburgh, EH9 3FD, UK

$^{\nabla\nabla\nabla\nabla}$ School of Chemical Sciences $\&$ Engineering, Yachay Tech University, 100119- Urcuqu\'i, Ecuador

$^{\dagger\dagger\dagger\dagger}$ Department of Chemistry, University of Pittsburgh, Pittsburgh, PA, USA, 15260

$^{\ddagger\ddagger\ddagger\ddagger}$ Center for Soft Nanoscience, University of Münster, 48149 Münster, Germany

\bigskip

$^{\ast}$ Authors to whom correspondence should be addressed

$^\equiv$ Juan José Sáenz passed away on March 22, 2020

\begin{abstract}
\par Chiral degrees of freedom occur in matter and in electromagnetic fields and constitute an area of research that is experiencing renewed interest driven by recent observations of the chiral-induced spin selectivity (CISS) effect in chiral molecules and engineered nanomaterials. The CISS effect underpins the fact that charge transport through nanoscopic chiral structures favors a particular electronic spin orientation, resulting in large room-temperature spin polarizations. Observations of the CISS effect suggest opportunities for spin control and for the design and fabrication of room-temperature quantum devices from the bottom up, with atomic-scale precision. Any technology that relies on optimal charge transport, including quantum devices for logic, sensing, and storage, may benefit from chiral quantum properties. These properties can be theoretically and experimentally investigated from a quantum information perspective, which is presently lacking. There are uncharted implications for the quantum sciences once chiral couplings can be engineered to control the storage, transduction, and manipulation of quantum information. This forward-looking perspective provides a survey of the experimental and theoretical fundamentals of chiral-influenced quantum effects, and presents a vision for their future roles in enabling room-temperature quantum technologies.

\end{abstract}



\section*{Overview}
\subsection*{Effects Involving Chiral Matter and/or Chiral Probes}

\indent \par Chiral matter (\textit{i.e.}, structures that can occur with left- or right-handed non-superimposable symmetry; equivalently, that lack improper rotation axes) offers unparalleled opportunities for the exquisite control of electron and spin transport 
due to its extraordinary optical, electronic, and magnetic properties. These properties are often observed at near-room temperatures, and suggest that quantum devices based on chiral matter could operate at similar high temperatures if properly designed. In this perspective, we stress the potential of chiral matter and fields, in particular with respect to their little explored applications in the quantum sciences. We detail how chiral matter presents advantages for the scalability and flexibility of molecular architectures interfaced with low-dimensional materials; 
the operation of quantum devices at room temperature and in noisy photonic, phononic, and electronic environments; and the advance of investigations into the emergent field of quantum biology.

\par In particular, the chiral-induced spin selectivity (CISS) effect is an unusual spin-valve-like behavior first observed in biological structures, and later replicated in technological applications, existing even in diamagnetic systems. The CISS effect is charge transport through nanoscopic chiral structures that favors a particular component of electron spin in the direction of propagation, an effect also referred to as `spin polarization,' all in the absence of external magnetic fields. 

\par The CISS effect can be physically understood as an electron scattering process in a molecular potential where spin-orbit (SO) interactions are strong enough and where both space inversion and time reversal symmetries are broken. This combination of physical constraints translates into spin polarization and spin filtering arising from the asymmetric spin- and chirality-dependent transmission probabilities. The CISS effect has been established in electron transfer, electron transport, and bond polarization through chiral centers or complex helical chiral structures. 

\begin{figure}[!ht]
  \includegraphics[width=5in]{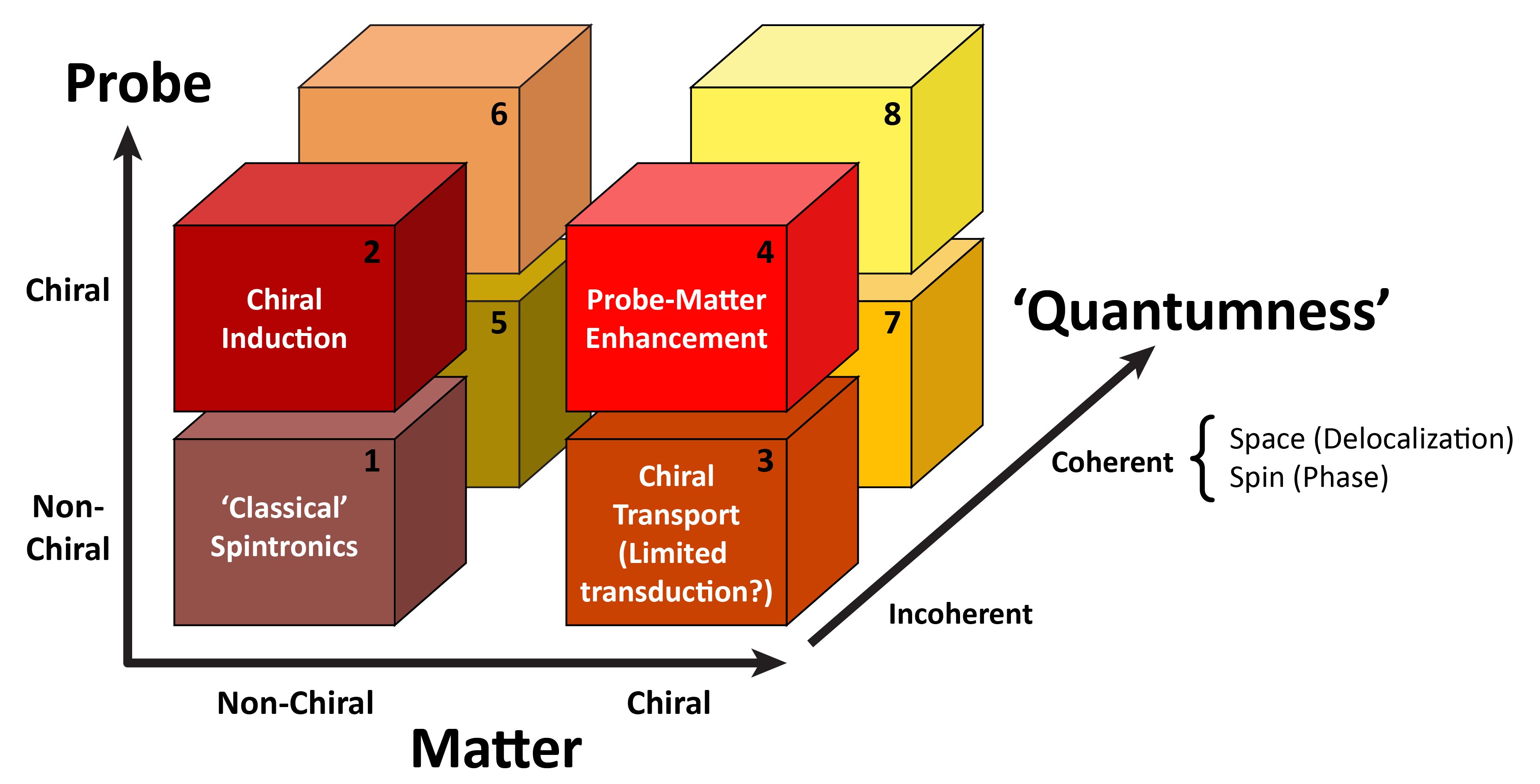}
  \caption{\textbf{The chiral experimenter's space.} The material (matter) under study and the `probe' (\textit{e.g.,} electromagnetic fields) can be either chiral or non-chiral. We expect different interaction strengths and rules to be valid in each `box'. A third  `quantumness' axis points to the fact that matter-probe interactions could be classical \textcolor{black}{(boxes 1-4)} or, in principle, preserve coherences \textcolor{black}{(boxes 5-8)}.}
  \label{Axes}
\end{figure}

With this article being a perspective and introducing the CISS effect for spatially chiral matter, it is exciting to ask what general degrees of freedom the field of chirality provides, especially in the context of quantum information processing. Indeed, there are three orthogonal axes that span across the emerging `chiral experimenter's universe' (Fig.\ \ref{Axes}) consisting of: (i) the structural type of chirality (\textit{i.e.}, matter); (ii) the presence of chirality in the interrogation or spin-inducing method (\textit{i.e.}, probe); and (iii) the presence of ‘quantumness’ during the matter-probe interactions. In their plurality, these three axes define a wide space that is yet to be fully investigated. 

Commenting on prominent octants of this chiral space, both a non-chiral matter and probe represents the field of classical spintronics \cite{jansen2003spin, bader2010spintronics}, where spin currents are induced into, typically, semiconductor--magnetic junctions leading to spin-valve devices such as transistors \cite{jansen2003spin} (box 1, Fig.\ \ref{Axes}). The effect of local circular dichroism (CD) in isotropic matter interacting with orbital angular momentum-carrying laser light has been theoretically predicted in non-chiral hydrogen-like media.\cite{afanasev2017circular} Based on violation of angular momentum selection rules in orbital angular momentum (OAM) photon--atom interactions \cite{afanasev2018experimental}, it has been also predicted that similar effects can be observed in trapped Ca$^{40}$ ions (box 2, Fig.\ \ref{Axes}).\cite{solyanik2019excitation} An experimental validation of orbital angular momentum-induced CD can be found in Ref.\ \citenum{zambrana2014angular}. This result shows that geometric structural chirality is not required for inducing handedness into matter. Nonetheless, future research shall investigate the limits of this approach such as the degree of obtainable transduction efficiency, or the possibility of achieving a coherent transduction of states and currents, thus preserving coherence indicated on the quantumness axis of Fig.\ \ref{Axes}. In other words, the scientific question of whether non-chiral materials can exhibit and preserve quantum coherence is not definitely answered yet and, certainly, its operational limits leave much room to be explored. Incidentally, here we define coherence as being present in either chiral domain as spin (\textit{i.e.}, phase), or spatial (\textit{i.e.}, delocalization).

Possible experimental implementations for the `matter' axis include transductions into molecules, nanostructures such as two-level systems delivered by quantum dots (QDs), and circuit-based metamaterials. These inspire nanofabrication of multipolar structures with spatial dimensions fitted to achieve non-trivial chiral response in photonics.\cite{capasso2018} This research direction also encompasses the magnetic response \cite{franke2017light}, optical devices such as those in passive integrated photonics or electro-optics \cite{amin2020sub}. Also of great interest are free-space reconfigurable optical systems such as spatial light modulators  used to realize optical coherent Ising machines \cite{pierangeli2020noise}, and digital mirror displays employed in convolutional neural networks.\cite{miscuglio2020massively} 

The aforementioned CISS effect and its early demonstrations\cite{campbell1985spin}) is represented by box 3 in Fig.\ \ref{Axes}, where a non-chiral stimulus, such as an electrical, optical or magnetic probe, induces a current exhibiting spin polarizations of different degrees. If both the probe and matter are chiral, we speculate there to be an enhanced interaction of probe and matter (box 4, Fig.\ \ref{Axes}). Here, we can ask what transduction efficiencies might one be able to obtain when only the probe or the matter is chiral (\textit{i.e.}, boxes 2 and 3 in Fig.\ \ref{Axes})? 

With this plethora of probe-matter interaction enhancement opportunities, the two main questions in the context of chirality remain: (i) What transduction efficiencies are possible?,  and (ii) To what extent can coherence be preserved, both during the transduction, and also while propagating once induced? The ability to preserve long-range coherences will determine the number of correlated spin-quantum states which translates into scale-up opportunities for quantum information processors. \textcolor{black}{Thus, the remaining four quadrants of Fig. \ref{Axes} (boxes 5-8) convey the same combinations of chiral and non-chiral probes and matter but with the addition of coherence, which can encompass spatial coherence or spin coherence.}


\par In technological quantum devices, achieving spin polarization requires a sophisticated degree of `quantum control' \emph{via} engineered electromagnetic excitation. Nature seems to have found its own way towards spin polarization. As a consequence of the CISS effect, chiral isomers (\textit{i.e.}, enantiomers) have opposite electron spin orientation preferences, which might be explored by the pharmaceutical industry. The CISS effect might also have important biological implications, as proteins and many biomolecules, including DNA and amino acids, are chiral. Finally, it was recently demonstrated that electron transport through bacteria nanowires is spin polarized.\cite{Mishra2019} If spin is harnessed in cellular electron transport processes, quantum information transfer in living environments might be possible. In addition, electron transfer through chiral structures is more efficient than that through comparable achiral structures, and under some conditions can be temperature independent, which are critical  to technologies relying on optimal charge transport (\textit{i.e.}, the entire quantum device industry). Finally, the complex response of chiral molecules to electromagnetic fields, involving both the electric and magnetic polarizabilities, suggests the potential of controlling and manipulating interactions and the imprinting of chiral characteristics onto photons, phonons and materials.
 
\par Unveiling the quantum mechanisms behind electron spin transport \emph{via} the CISS effect may enable the manipulation of the effect for technological and theranostic advantage -- especially within the realm of the quantum sciences -- as we aim to illustrate with this review.

\subsection*{Outline of this Review}
\indent \par We review theoretical and experimental aspects of quantum effects in chiral materials as follows:

\begin{itemize}
\item Chirality and Spin Measurements: Experimental measurements demonstrating the role of spin, chirality, and the CISS effect;  
\item Chirality and Spin Theory: Theoretical predictions and modelling tools that inform the understanding of CISS;
\item Engineered Chiral Materials: Materials design strategies that are being developed to engineer and tune chiral properties;
\item Chirality in Biology: Potential chirality-mediated effects in biology, and predict potential future applications in the field;
\item Chiral Degrees of Freedom in the Interaction of Matter and Electromagnetic Fields: Potential avenues to enhance interactions between matter and electromagnetic fields \emph{via} chiral degrees of freedom; 
\item Chirality in the Quantum Sciences: Forward-looking applications of the CISS effect in the quantum sciences.
\item \textcolor{black}{Future Outlook and Conclusions: Summary of main points of this paper and perspectives on future directions for the field}
\end{itemize}

\par This perspective attests to the potential for harnessing chirality and CISS as tools in fields as diverse as quantum information science, spintronics, nanotechnology, and control of biological systems at the nanoscale. Importantly, these advances in our understanding and observations of chirality-based quantum phenomena are poised to be incorporated into spin-based quantum technologies that may function at elevated temperatures.


\section*{Chirality and Spin Measurements}\label{measurements}

\subsection*{Electrochemistry Experiments}\label{echem}

\indent \par
\begin{figure}[!ht]
  \includegraphics[width=6.5in]{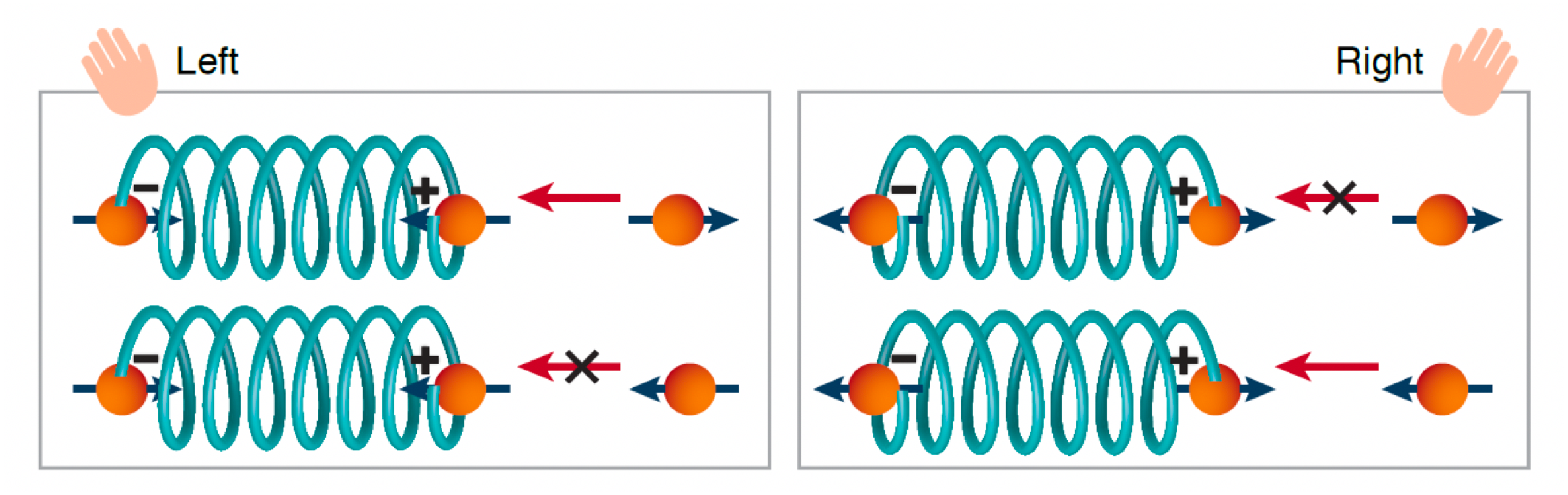}
  \caption{\textbf{Illustration of a spin-polarization-induced enantioselective reaction mechanism.} When an electron approaches a chiral molecule, charge rearrangement occurs and the molecule becomes charge polarized with the electron attracted to the positive pole of the molecule. Depending upon the molecule’s handedness (left-handed or right-handed) and the spin orientation of the electron, the interaction is more favored or less favored for a given spin polarization. Reproduced with permission from Ref.~\citenum{Bloom2020}. Copyright 2020 by the Royal Society of Chemistry.}
  \label{FigElectrochemistry1}
\end{figure}

 Electrochemical processes add to or remove electrons from chemical systems. \textbf{Fig.\ \ref{FigElectrochemistry1}} shows that redox processes in chiral molecules can be spin selective. When an electron approaches a chiral molecule, it induces charge polarization. An induced electric dipole is formed and the electron is attracted to the positive pole. It was found that in chiral molecules, charge polarization is accompanied by spin polarization due to the CISS effect;\cite{Kumar2017,Ziv2019} which spin is associated with which electric pole depends on the handedness of the molecule. Therefore, when spin-polarized electrons approach a chiral system,\cite{Ziv2019} their ability to attach to the molecule is higher if the spin on the positive pole is polarized opposite to the spin of the approaching system. This preference results in enantioselectivity given that the electron is spin-polarized.\cite{Bloom2020}

\par
\begin{figure}[ht!]
  \includegraphics[width=5in]{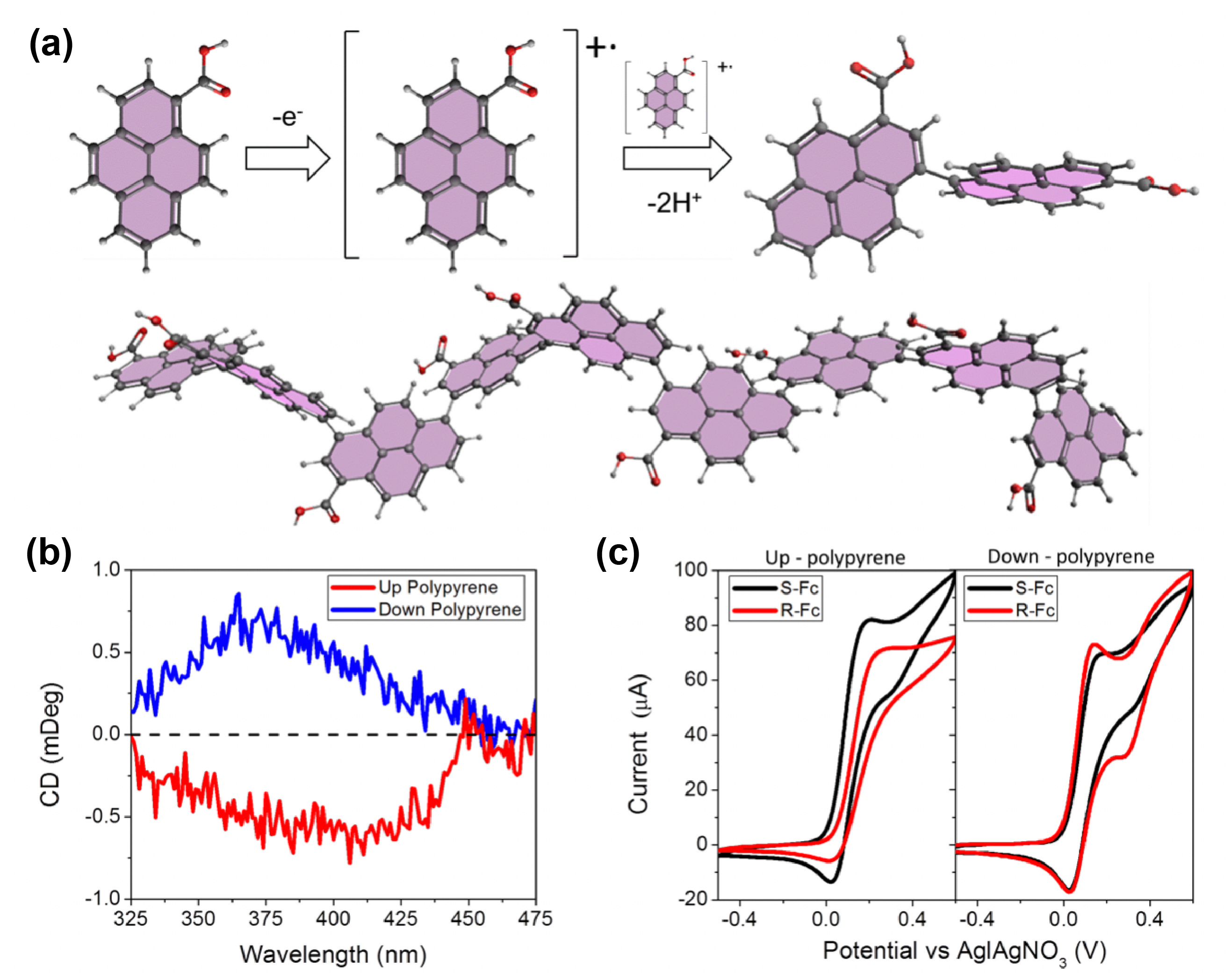}
  \caption{\textbf{Reaction scheme and chirality in electropolymerized polypyrene.} \textbf{(a)} Reaction scheme for the polymerization of 1-pyrenecarboxcylic acids into polypyrene which exhibits a helical twist (see main text for more details). \textbf{(b)} Circular dichroism spectra for electrodes coated with polypyrene where a magnetic field was applied Up (red) or Down (blue) during electropolymerization. \textbf{(c)}  Electrochemistry measurements on (S)- (black) or (R)-ferrocene (red) with the Up (left) or Down (right) polypyrene-coated working electrodes. Reproduced with permission from Ref.~\citenum{Metzger2020}. Copyright 2020 by John Wiley and Sons.}
  \label{FigElectrochemistry2}
\end{figure}

 Recently, it was found that spin selectivity can control both electrochemical reduction and oxidation.\cite{Metzger2020} As an example of an oxidation process, electropolymerization of 1-pyrenecarboxylic acid was performed on a magnetic electrode (10 nm of Ni and 10 nm of Au on ITO) that was magnetized either `up' or `down' relative to the electrode surface.  \textbf{Fig.\ \ref{FigElectrochemistry2}a} shows a reaction scheme for the formation of polypyrene. Initiation of the reaction involves electro-oxidation of the monomer unit to form a radical cation. The steric constraints of the pyrene rings lead to a propeller-like arrangement of the monomers; control over their stereo arrangement imparts axial chirality into the polymer chain. \textbf{Fig.\ \ref{FigElectrochemistry2}b} shows the CD spectra of the pyrene polymer films on the electrode surface. In turn, the red curve shows the CD spectrum with the electrode magnetized in the `up' direction and the blue curve corresponds to the case for magnetization in the `down' direction. The red and blue curves exhibit opposite Cotton effects in pyrene's excimer spectral region. 

\par The chirality of the polymer-coated electrode was confirmed by performing cyclic voltammetry (CV) with a chiral ferrocene (Fc) redox couple.  \textbf{Fig.\ \ref{FigElectrochemistry2}c} shows voltammetry data collected using the polypyrene-coated films as working electrodes for two different enantiomerically pure solutions of chiral ferrocene: ($S$)-Fc (black) and ($R$)-Fc (red). It is evident from the voltammetric peak currents that the `up' grown electrode is more sensitive to the ($S$)-Fc, while the `down' grown electrode is more sensitive to the ($R$)-Fc. Similar dependencies for redox properties with chiral working electrodes were reported elsewhere, and further corroborate the chirality demonstrated in the circular dichroism measurements.\cite{Mogi2007} This result demonstrates how chiral spin transport can lead to highly amplified downstream chemical products.

\par Spin-related effects have also been observed in cyclic voltammetry (CV) experiments performed under applied magnetic fields on a non-ferromagnetic electrode modified with a thin electroactive oligothiophene film.\cite{Benincori2019} When flipping the magnet's up/down magnetic moment orientation, the CV peaks of two achiral, chemically reversible Fe(III)/Fe(II) redox couples in aqueous or organic solution undergo impressive potential shifts (up to nearly 0.5 V depending on protocol conditions), by changing the film's ($R$)- or ($S$)-configuration. This observed effect is another manifestation of the spin-chirality relation in electrochemical processes.

\subsection*{Enantioseparation Experiments}

\indent \par
\begin{figure}[!ht]
  \includegraphics[width=4in]{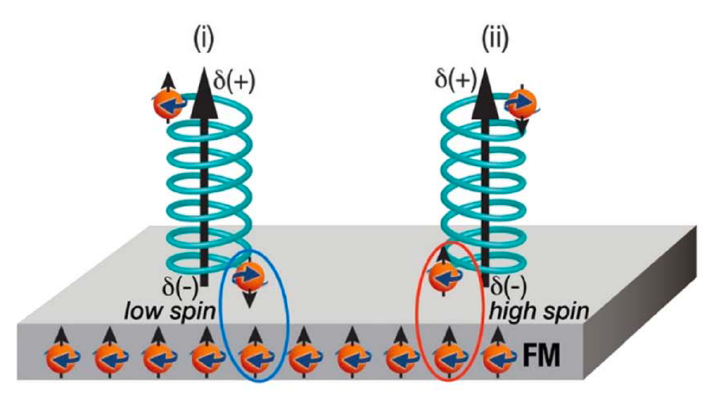} 
  \caption{\textbf{Chiral-induced spin selectivity effect and ferromagnetic substrates}. As a chiral molecule approaches the ferromagnetic (FM) substrate, its charge polarization generates a spin polarization at the two ends of the molecule. For a specific enantiomer, the interaction between the magnetized surface and the molecule (circled in blue and red) follows a low-spin or a high-spin potential, depending on the direction of magnetization of the substrate. Reproduced with permission from Ref.~\citenum{BanerjeeGhosh2018}. Copyright 2018 by The American Association for the Advancement of Science.}
  \label{FigEnantioseparation}
\end{figure}

 The relation between spin and chirality \emph{via} the CISS phenomenon naturally leads one to consider the possible interactions between chiral molecules and ferromagnetic surfaces.\cite{Naaman2019, Tassinari2019, Dor2017, BanerjeeGhosh2018, Naaman2020} This interaction should be spin sensitive due to short-range magnetic exchange interactions: as chiral molecules approach the surface, charge reorganization and spin polarization should take place, depending on the handedness of the molecules. We refer below to two specific examples of this mechanism:

\textbf{i)} It was recently demonstrated that magnetization switching of ferromagnetic thin layers can be induced solely by the adsorption of chiral molecules without external magnetic fields or spin-polarized currents.\cite{Dor2017} The mechanism for chiral molecules to induce magnetization switching is shown in \textbf{Fig.\ \ref{FigEnantioseparation}}.  The effect of adsorbed chiral molecules on the properties of a ferromagnetic substrate was examined by studying the adsorption of \textsc{l}- and \textsc{d}-oligopeptides on thin ferromagnetic films, with an initial magnetization that was not defined.  The direction of the magnetization was found to depend on the handedness of the adsorbed chiral molecules. Importantly, fewer than 10$^{13}$ electrons per cm$^2$ are sufficient to induce a reversal of the magnetization on the ferromagnetic layer in the direction perpendicular to the surface. We note that the current density required for common mechanisms in modern magnetoresistive random-access memory, such as spin-transfer torque memories, is 10$^{25}$ electrons per cm$^2$). The high efficiency of magnetization results from the molecule--substrate exchange interaction. As such, this concept could be used to achieve simple surface spintronic logic devices.

\textbf{ii)} It was shown that enantiomers can be separated by adsorbing them on a ferromagnetic substrate that has a magnetization perpendicular to the surface.\cite{BanerjeeGhosh2018} Unambiguous enantioselectivity on a ferromagnetic substrate was obtained for a variety of chiral molecules and magnetic substrates.  That is, while one enantiomer adsorbs more rapidly than the other when the magnetic dipole is pointing up, the other adsorbs faster  when the substrate is magnetized in the opposite direction. The interaction between the chiral molecules and the magnetized substrate is not affected by the magnetic field but by the interaction between the spin- polarized molecule and the spin of the electrons  on the substrate. As above, the effect is associated with the magnetic-exchange interaction of the spin-polarized molecules with the spin-polarized substrate. This phenomenon creates prospects for generic approaches to enantiomeric separations, which are critically important to the chemical industry.


\subsection*{Scanning Probe Microscopy (SPM) Experiments}\label{spmic}

\indent \par Direct methods to study spin-selective conductive properties of self-assembled chiral molecular monolayers (SAMs) using scanning probes were used with great success over the past 10 years. In an experimental demonstration of this methodology, Xie \textit{et al}.\ used conductive atomic force microscopy (c-AFM) to measure room temperature magnetization-dependent conductivity through Au-nanoparticle (AuNP)-terminated SAMs of double-stranded DNA (ds-DNA) on thin-film Ni.\cite{Xie2011} Although the variance between individual measurements is significant, likely due to local variations in the electrical contact between the AFM tip and the Au NP, comparison of current-voltage traces collected under these varying conditions indicates asymmetry in the Ni-DNA-AuNP junction conductivity, depending on the magnetization orientation of the Ni substrate, thereby revealing a preference for electrons with their spin aligned parallel to their momentum in tunneling transport through right-handed dsDNA. Control measurements conducted with Au instead of Ni thin films confirmed that the conduction asymmetry depends on the presence of both a source and a sink of spin-polarized electrons (the magnetic thin film), and a spin-discriminating conductive element (the ds-DNA). This behavior contrasts starkly with that of standard organic spin valves, in which organic layers typically play the role of weakly interacting non-magnetic spacing layers, carrying spin-polarized electrons from one magnetized material to another.\cite{Xiong2004, Schmaus2011} In contrast, the chiral ds-DNA layer behaves as an active element, more similar to a traditional inorganic spin valve. Mishra \textit{et al}.\ recently used a similar approach to study length-dependent spin selectivity in conduction through SAMs of ds-DNA and $\alpha$-helical peptides.\cite{Mishra2020a}

Although the application of SAM-based surface modification methods to reactive metal thin films (such as Ni) has been demonstrated,\cite{Cheung2020} the formation of SAMs on reactive surfaces is known to be complicated by the competitive formation of surface oxide layers, which may displace molecules, limit molecular adsorption, and worsen film uniformity.\cite{Mekhalif1997, Hohman2011} In the context of chiral spintronics, this challenge has traditionally been overcome by using a thin inert capping layer, such as Au, to protect the magnetic layer from oxidation.\cite{Abendroth2017, Abendroth2019, Stemer2020} However, the capping layer strategy must be carefully implemented; thicker capping layers of Au, which as a heavy element may induce strong SO coupling effects on conducted electrons, have been shown to lead to spin randomization and nullification of spin-dependent effects in charge transport through the chiral molecules assembled thereupon.\cite{Mondal2015, Ghosh2020} The capacity to apply conductive atomic force microscopy (c-AFM) to characterize conduction in chiral molecular films without the need for a capping layer, as exhibited by magnetic c-AFM (mc-AFM), is therefore highly desirable.

Bloom and colleagues used mc-AFM to study spin-selective conduction through chiral CdSe QDs drop cast onto highly ordered pyrolytic graphite.\cite{Bloom2016} When functionalized with chiral ligands, QDs show strong chiro-optical activity as a result of orbital hybridization between the highest occupied molecular orbitals of the chiral ligand and the valence band states of the QDs.\cite{Choi2016} In measurements analogous to those conducted by Xie \textit{et al}., Bloom and colleagues characterized the conductivity of hybrid QD-chiral molecule junctions as a function of Co-Cr mc-AFM tip magnetization orientation. Once again, a distinct asymmetry is apparent when comparing current-voltage traces collected for QDs functionalized with \textsc{l}- \textit{vs.} \textsc{d}-cysteine ligands. Tip magnetization persistence was confirmed by comparing magnetic force microscopy images before and after sample analysis. Analogous experiments with spin-polarized and spin-sensitive scanning tunneling microscopy (STM) remain an opportunity for the field.\cite{Bonnell2012}

\begin{figure}[!ht]
  \includegraphics[width=5in]{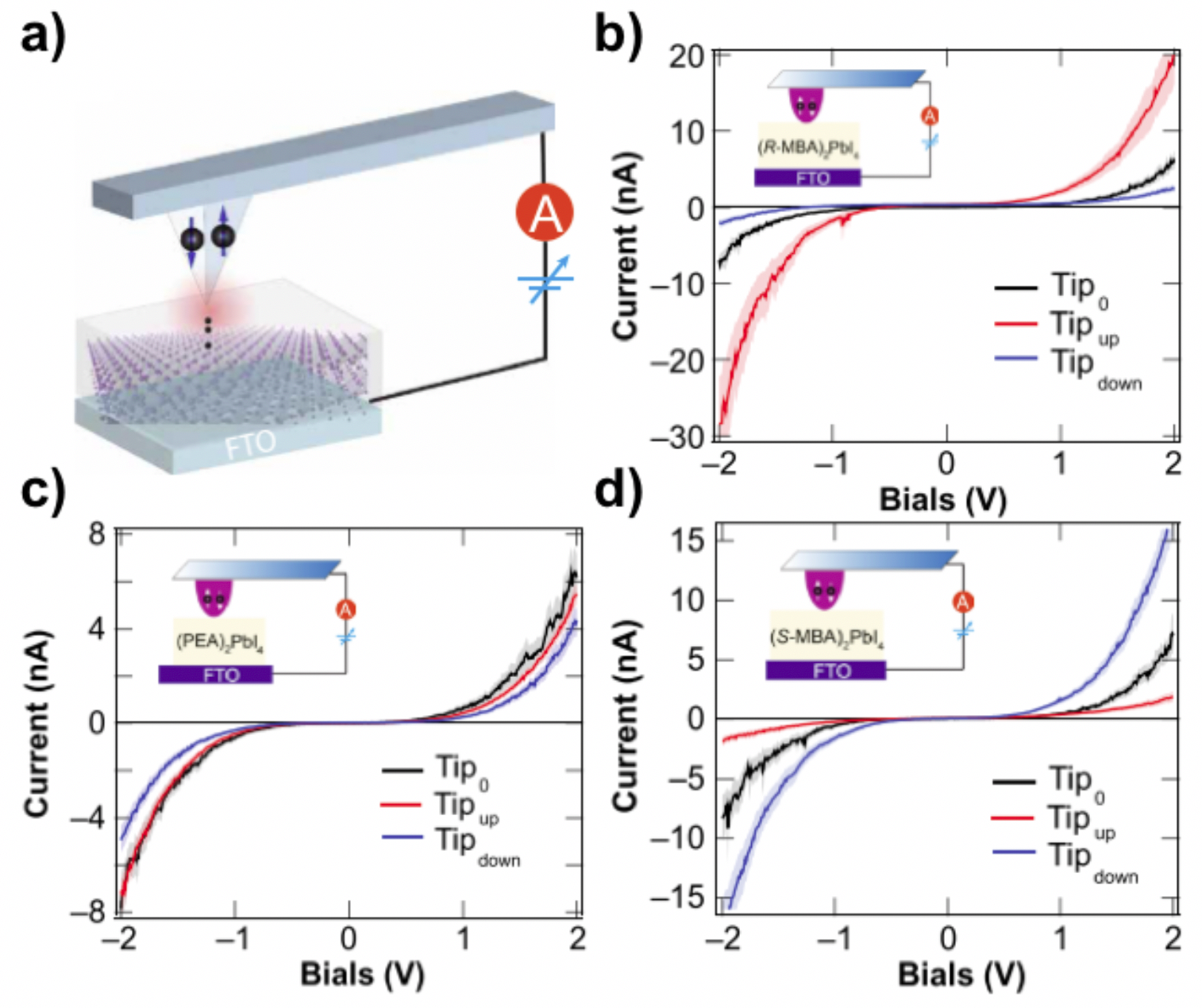} 
  \caption{\textbf{(a)} Experimental scheme for magnetic conductive atomic force microscopy (AFM) measurement of chiral perovskite thin films. \textbf{(b)-(d)} Current--voltage traces collected for thin films of lead--iodide perovskites containing chiral R-methylbenzylammonium, achiral phenylmethylamine, and chiral S-methylbenzylammonium, respectively, as a function of tip-magnetization orientation. Reproduced with permission from Ref.~\citenum{Lu2019}. Copyright 2019 by the American Association for the Advancement of Science.}
  \label{FigScanningProbe}
\end{figure}

Subsequent applications of mc-AFM have moved beyond chiral QDs toward characterizing spin-selective conductivity through thicker films of ordered chiral materials. Lu \textit{et al}.\ reported conduction asymmetries of up to 86\% (measured as percent difference in current at 2 V for `tip-up' and `tip-down' magnetization conditions) in solution processed 50 nm thin films of chiral lead--iodide hybrid perovskites, as shown in \textbf{Fig.\ \ref{FigScanningProbe}}.\cite{Lu2019} Similar values of spin-selectivity were reported recently in c-AFM measurements of conduction through supramolecular chiral nanofibers assembled on Ni thin films capped with gold.\cite{Kulkarni2020} By tuning the chirality (R- or S-) of the organic methylbenzylammonium constituent, Lu and colleagues showed control over the handedness of the perovskite films studied. More recent reports focused on similar tin--iodide perovskites yielded even higher spin-selectivity values,\cite{Lu2020} nearing 94\% and underscoring the utility of conductive probe microscopy as a powerful tool for rapid and direct characterization of electronic phenomena in chiral materials, particularly as the materials continue to grow in relevance to the broader spintronics community. 

\subsection*{Molecular Electronics Experiments}

\indent \par Naaman and co-workers reported  a large asymmetry in the transmission probabilities of polarized electrons by thin films of chiral molecules.\cite{Ray1999}  This was supported by electron transmission measured through chiral monolayers with STM, demonstrating that chiral systems can act as spin filters and spin polarizers,\cite{Aragones2017, Varade2018} allowing the preferential transmission of only one spin component. Theory indicates that this is possible due to enhanced SO interactions and space and time reversal symmetry breaking, which are the result of transport through chiral symmetries and an electrochemical gradient induced by an external voltage or a free energy gradient \cite{Yeganeh2009, Guo2012}. 

\par Break-junction devices provide an alternative class of measurements that use nano-\linebreak structured, moveable electrodes to make contact with single molecules to study the charge transport properties \cite{Xu2003, Smit2002, Venkataraman2006, Martin2008}. In this system, two electrodes are brought into and out of contact with the molecules of interest. As the two electrodes are withdrawn, the current is measured. When a molecule binds between the two electrodes, a plateau is observed in the current \textit{vs.}\ distance trace. By measuring thousands of these traces, it is possible to statistically determine the conductance of a single-molecule junction. These systems were used extensively to study charge transport in chiral molecules such as DNA,\cite{Xu2004, Hihath2005, Guo2016, Liu2013, Kawai2010} RNA,\cite{Li2018, Li2016} peptide nucleic acid (PNA),\cite{Paul2010}  peptides,\cite{Sek2006, Xiao2004} and proteins,\cite{Artes2012, Zhang2020} and were used to explore spin-selectivity in a variety of  molecules by either applying a magnetic field or by using a spin-polarized electrode to inject polarized spins.\cite{Aragones2016, Osorio2010}
	
Recently, these approaches were combined to explore spin selectivity in a diamagnetic, helical peptide chain that formed an $\alpha$-helix.\cite{Aragones2017} These experiments allow measurement of the spin polarization power of a chiral molecule and provide direct evidence of the spin-filtering capabilities of these systems. By changing both the chirality of the molecule (by using both \textsc{l}- and \textsc{d}- isomers) and the magnetization orientation of the injecting electrode, a clear change in the charge transport properties of the molecular system through these four possible conditions was demonstrated, besides a spin polarization power (capability to spin‐polarize electrical current) of ~60\%.

Moving forward, nucleic acid systems may provide a fruitful framework for understanding CISS behavior at the single-molecule level, as there is a large variety of conductance measurements on these molecules.\cite{Xu2004, Hihath2005, Guo2016, Liu2013, Kawai2010} Synthetic approaches will allow ready control of both the length and the helical pitch of the molecules and, although there can be some structural changes when DNA is dehydrated,\cite{Dulic2009, Artes2015, Adessi2003} the helices can survive both vacuum conditions and wide temperature ranges.\cite{Kasumov2004}

The conductive probe and spin-exchange methods used to study spin-dependent transport or charge redistribution involve multiple contacts or molecules within an interface, which may convolute spin-selective tunneling effects across individual chiral molecule bridges. Alternatively, scanning tunneling microscopy break junction (STM-BJ) techniques provide a measurement of conductance through single molecules, enabling hundreds to thousands of measurements that provide statistically relevant analyses.\cite{Aragones2017} Briefly, in a STM-BJ experiment, the STM tip is driven into and out of contact to/from the sample of interest.

Previous works have measured the conductance of single-peptide strands of sequences with strong $\alpha$-helical propensity.\cite{Sek2006, Sek} The relatively low registered conductance levels, \textit{i.e.}, 10$^{-7}$--10$^{-5}$G$_0$ (G$_0$ = 77.4 $\mu$S), for helices of 2 to 3 nm in length, and the observed fairly high length decay penalty constant, \textit{i.e.}, 5 nm$^{-1}$, suggest an off-resonant tunneling mechanism as the dominant charge transport path in these molecular systems. The latter is supported by the recently reported temperature-independent behavior of charge transport observed in protein systems based on $\alpha$-helical moieties.\cite{ruiz, ron} 

\begin{figure}[!ht]
  \includegraphics[width=5.5in]{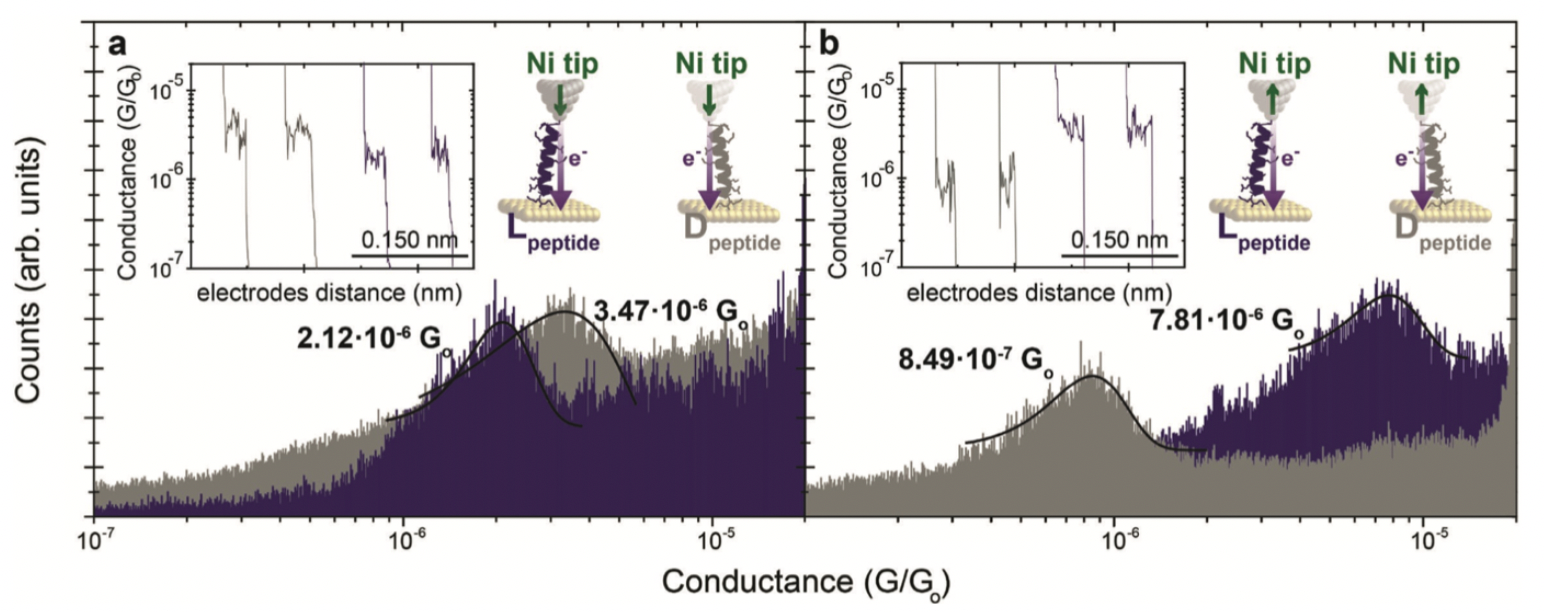} 
  \caption{Conductance histograms for hundreds of single-peptide junctions collected in different STM-BJ experiments with Ni tips magnetized \textbf{a)} down and \textbf{b)} up for both left- and  right-handed alpha-helical peptides. These peptides are composed of 22 amino acid residues bridges attached to gold substrates. Insets depict representative current \textit{versus} pulling traces with well-defined single-molecule plateau features. Conductance values were extracted from Gaussian fits to the histograms. Adapted with permission from Ref.~\citenum{Aragones2017}. Copyright 2017 by John Wiley and Sons.
}
  \label{FigSTMBJ}
\end{figure}

More recently, using a spin-polarized STM-BJ method, Aragon\`es \textit{et al}.\ found that conductance between ferromagnetic Ni tips and Au surfaces bridged by sulfur-terminated $\alpha$-helical peptides (cysteine residues) depends on the tip magnetization direction and handedness (L or D) of amino acid residues,\cite{Aragones2017} as shown in \textbf{Fig.~\ref{FigSTMBJ}}. When a spin-polarized source of electrons is used to inject charge into one of the $\alpha$-helical barrel isomers (\textsc{l}- or \textsc{d}-), the conductance values are observed to consistently depend on both the particular orientation of the spin injected and handedness of the chiral helical peptide during the charge transport measurement. The spin polarization power of a single peptide strand can be then calculated by the conductance relation (G$_{L/D}\uparrow$ - G$_{L/D}\downarrow$)/(G$_{L/D}\uparrow$ + G$_{L/D}\downarrow$), being GL/D the conductance of either the \textsc{l}- or \textsc{d}-isomer and the arrows indicated the orientation of the injected electron spin. Exceedingly high spin polarization powers scoring 60\% are registered for a single $\alpha$-helical peptide barrel of 22 amino acids ($\sim$3 nm long). The results highlight a direct correlation between electron spin polarization and transport. Still, experimental evidence is lacking to unambiguously distinguish incoherent hopping from coherent tunneling in spin-dependent conductance through chiral molecules. Moving forward, systematic investigations of the influence of monomer sequence, length, and molecular dipole on conductance (using spin-polarized STM-BJ techniques) could be employed to explore spin filtering that arises from the CISS effect.\cite{Xiang2015}

\subsection*{Spin Exchange Microscopy Experiments}\label{semic}

\indent \par Another scanning probe method that relies on non-ferromagnetic tips functionalized with chiral molecules has enabled local magnetic imaging similar to magnetic exchange force microscopy.\cite{Kaiser2007} Ziv \textit{et al}.\ found that transient spin polarization accompanying charge redistribution due to the CISS effect in the chiral molecules enables spin exchange interactions with magnetized samples, distinguishing domains magnetized up \textit{vs.} down by different forces exerted on the AFM cantilevers near the sample surfaces.\cite{Ziv2019} The forces were hypothesized to result either from symmetric or anti-symmetric spin alignment in the wave function overlap between molecules on the tips and the magnetized sample. Similarly, recent Kelvin-probe force microscopy measurements by Ghosh \textit{et al}.\ on ferromagnetic films coated with chiral SAMs  revealed electron spin-dependent charge penetration across the molecular interface.\cite{Ghosh2020} This dependence of wave function overlap between magnetized materials and chiral molecules on the spin-exchange interaction could be exploited for proximity-induced magnetization\cite{Dor2017} and enantiomer separations\cite{BanerjeeGhosh2018} that occur upon molecular adsorption, or to rationalize the spin-selective contributions to stereoselective interactions between biomolecules that result from induced dipole--dipole interactions.\cite{Kumar2017}
 
\subsection*{Experiments Involving Superconductivity}

\indent \par Spin-selective transport through chiral molecules shows unusual conduction phenomena near superconducting interfaces. In particular, chiral $\alpha$-helical polyaniline molecules that are in proximity to niobium superconductors (through chemisorption on metallic layers) show unconventional superconductivity. Shapira \textit{et al}.\ used scanning tunneling spectroscopy (STS) to analyze a multilayered polyaniline--gold bilayer-niobium substrate and found zero-bias conductance peaks embedded inside a bandgap from the tunneling spectra \cite{Shapira2018}. This zero-bias peak is reduced in the presence of a magnetic field but does not split, indicating an unconventional order parameter that is consistent with equal-spin triplet pairing \textit{p}-wave symmetry. In comparison, the adsorption of non-helical chiral cysteine molecules showed no change in the order parameter. 

\begin{figure}[ht!]
  \includegraphics[width=6in]{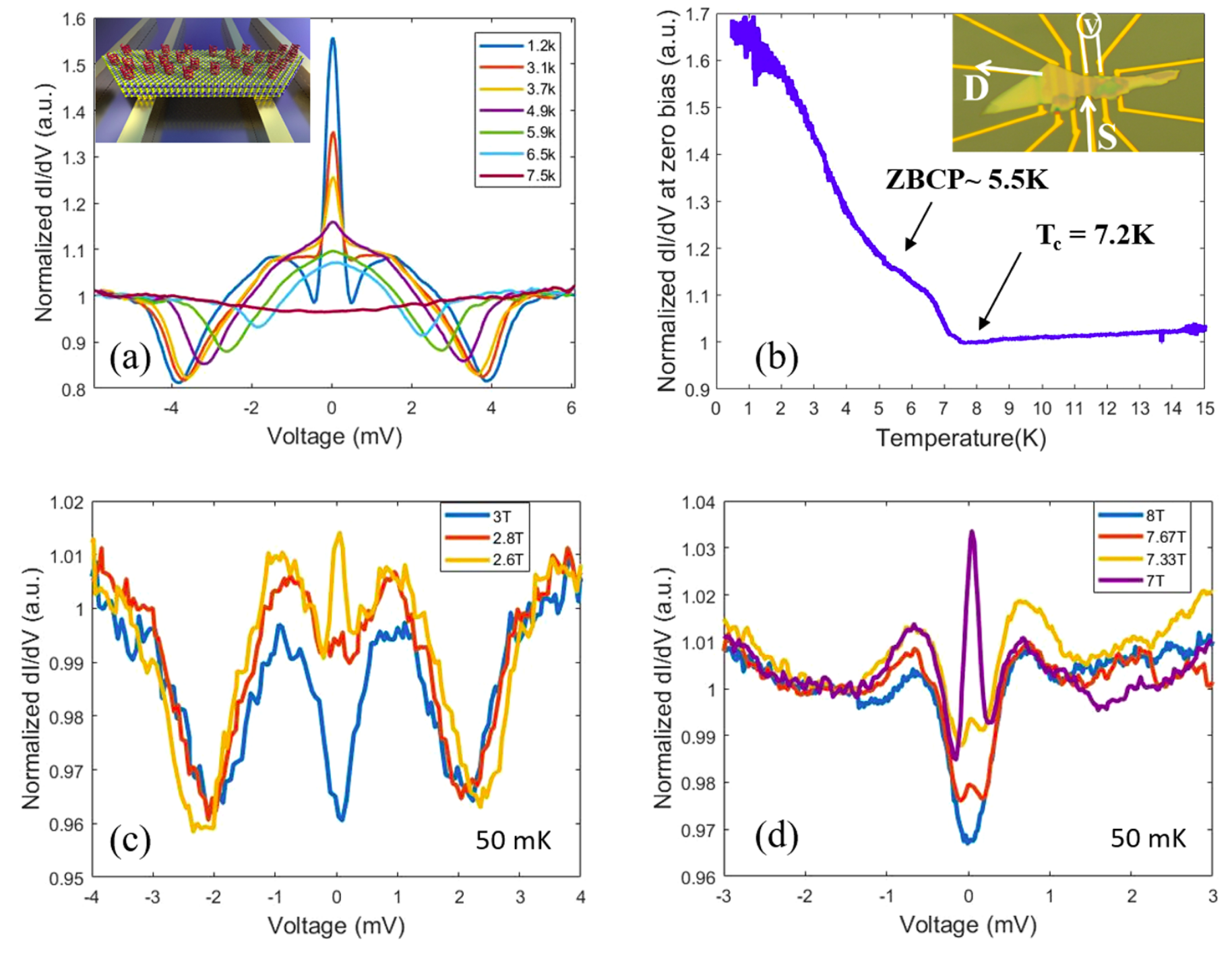} 
  \caption{\textbf{Conductance properties of a low-resistance (120 $\Omega$) NbSe$_2$/Au junction after chemisorption of chiral molecules on the NbSe$_2$ flake ($\sim$25 nm thick).} \textbf{(a)} Temperature dependence of dI/dV \textit{vs.} V spectra showing a distinct a zero bias conductance peak that vanishes at higher temperatures (but still below T$_\textrm{c}$). Inset: Illustration of a chiral-molecules/NbSe$_2$-flake/Au sample. \textbf{(b)} Temperature dependence of the conductance at zero bias with two transition temperatures marked by arrows: T$_\textrm{c}$ $=$ 7.2 K, where the zero bias conductance starts to rise due to the Andreev dome and 5.5 K where a zero bias conductance peak starts to appear. Inset: Optical image of the sample with the measurement scheme depicted. \textbf{(c,d)} Perpendicular \textbf{(c)} and parallel \textbf{(d)} magnetic field dependencies of the conductance spectrum, showing that in high magnetic fields, yet below the parallel and perpendicular critical fields (H$_\textrm{c$_2$}$) of bulk NbSe$_2$, the a zero bias conductance peak vanishes, revealing an underlying gap. Reproduced with permission from Ref.~\citenum{Alpern2019}. Copyright 2019 by the American Chemical Society.}
  \label{FigSuperconductivity}
\end{figure}

\par Junctions of superconducting niobium and metallic films bridged by chiral $\alpha$\nobreakdash-helical polyaniline induced \textit{p}-wave order-parameters, demonstrating phase-coherent transport through chiral organic films \cite{Sukenik2018}. Alpern \textit{et al}.\ analyzed proximity effects of chiral films in multilayered superconducting systems and found that chemically adsorbed $\alpha$\nobreakdash-helical polyalanine can induce magnetic-like state behavior. Specifically, in-gap states were found that were nearly symmetrical around a zero-conduction bias that closely resembled Yu--Shiba--Rusinov states \cite{Alpern2019}, see \textbf{Fig.\ \ref{FigSuperconductivity}}. 

\par Collectively, the observations described above provide evidence that chiral molecules can behave as magnetic impurities when in proximity to superconductors, enabling a wide range of potential applications for spin-selective transport through superconducting films.  

\section*{Chirality and Spin Theory} \label{theory}

\indent \par The CISS effect has been observed in a range of experiments and for a wide range of chiral molecules and materials. The CISS effect appears to be a room-temperature magnetic response due to internal molecular fields generated by electron SO interaction in chiral systems. This effect survives the inclusion of many-electron interactions and can coexist with other magnetic responses, including triplet radical formation, interstate crossing, and singlet fission, which provides fertile ground for spin manipulation. The CISS effect is closely related to exchange interactions, which play a central role in molecular recognition and chirality-induced effects on magnetic surfaces. 

\par The SO coupling that provides a source of magnetic fields for electrons atoms is a relativistic effect and is thus weak, namely on the order of a few meV, for the light-atom chiral molecules studied so far. However, it is sufficiently strong to generate a sizable spin polarization  through cumulative interactions with the chiral environment. For example, for transport through a large  molecule, an electron will encounter and transit many atoms. At each encounter, the SO interaction depends on the orbital orientation and will lead to some weak spin polarization. For a chiral molecule such as DNA, the electron spin polarization is cumulatively enhanced by the preferred orbital orientations of the many surrounding atoms as it travels through the molecule, leading to the CISS effect.\cite{Ray1999}

\subsection*{First-Principles Simulations with Chiral Degrees of Freedom}

\indent \par The vast majority of theoretical CISS-related studies rely on model Hamiltonians. Conventional electronic structure methods, such as density-functional theory (DFT), have only been applied in a few cases.\cite{Maslyuk2018, Dianat2020, Zoellner2020, Zoellner2020b} The DFT-based spin-dependent transport calculations, with the Landauer approach and including SO coupling, demonstrate the influence of helical structure on spin polarization \cite{Maslyuk2018, Dianat2020} and they correctly describe the increase of spin polarization with molecular length  as observed experimentally.\cite{Zoellner2020b, Mishra2020a}. However, the spin polarizations obtained using DFT are much smaller than the measured values, suggesting that the theoretical description may be missing some key ingredients. 

\par Recent experimental \cite{BanerjeeGhosh2018, Ziv2019} and theoretical \cite{Dianat2020, Zoellner2020b, Fransson2019} studies suggest that electronic exchange interactions may play a fundamental role in combination with the SO coupling \cite{Fransson2019}. In particular, exchange-related intermolecular interactions in arrays of helical molecules were shown to stabilize a broken spin symmetry (singlet) state, and the effect is absent in linear molecules \cite{Maslyuk2018, Dianat2020}. As a consequence, subsequent interactions with magnetized surfaces can facilitate symmetry breaking between enantiomers and may lead to enantiomer discrimination, which would be mediated by exchange interactions. Further indications of the relevance of exchange contributions was found in Ref.\ \citenum{Zoellner2020b}: changing the amount of Hartree-Fock exchange contribution exchange-correlation functionals strongly influences the size of the computed spin polarization in model helical molecular junctions. 

\par Another interesting finding from Ref.\ \citenum{Dianat2020} is the surprising ease with which formally closed-shell peptide helices can be spin-polarized in equilibrium when brought together in an organized fashion, as opposed to isolated helices. Interactions with metal surfaces were also found to be important. While past experience with DFT for spin-polarized molecules suggests caution \cite{Jacob2012}, these results point to intermolecular interactions and interfaces being important for the first-principles description of CISS. Likewise, the role of interfaces (possibly combined with the collective properties of molecular assemblies) for magnetic signatures in electron transport was explored experimentally.\cite{Li2020, Shi2017, Xie2016} In addition, going from a focus on a spin-polarizing helix to a full circuit analysis has been suggested \cite{yang2020detecting}. Beyond the description of exchange/electronic interactions, intermolecular and interface effects, it may be important to consider nuclear dynamics, as well as electronic.\cite{Guo2014,Wu2020} As a complement to efforts that aim to establish a comprehensive first-principles theory of CISS, chemical concepts extracted from first-principles calculations on helical molecules, such as electron transport pathways \cite{Tsuji2016, Garner2019} and imaginary components of the Hamiltonian \cite{Zoellner2020} may provide steps towards understanding structure--property relationships in CISS. 

\begin{figure}[!ht]
  \includegraphics[width=6in]{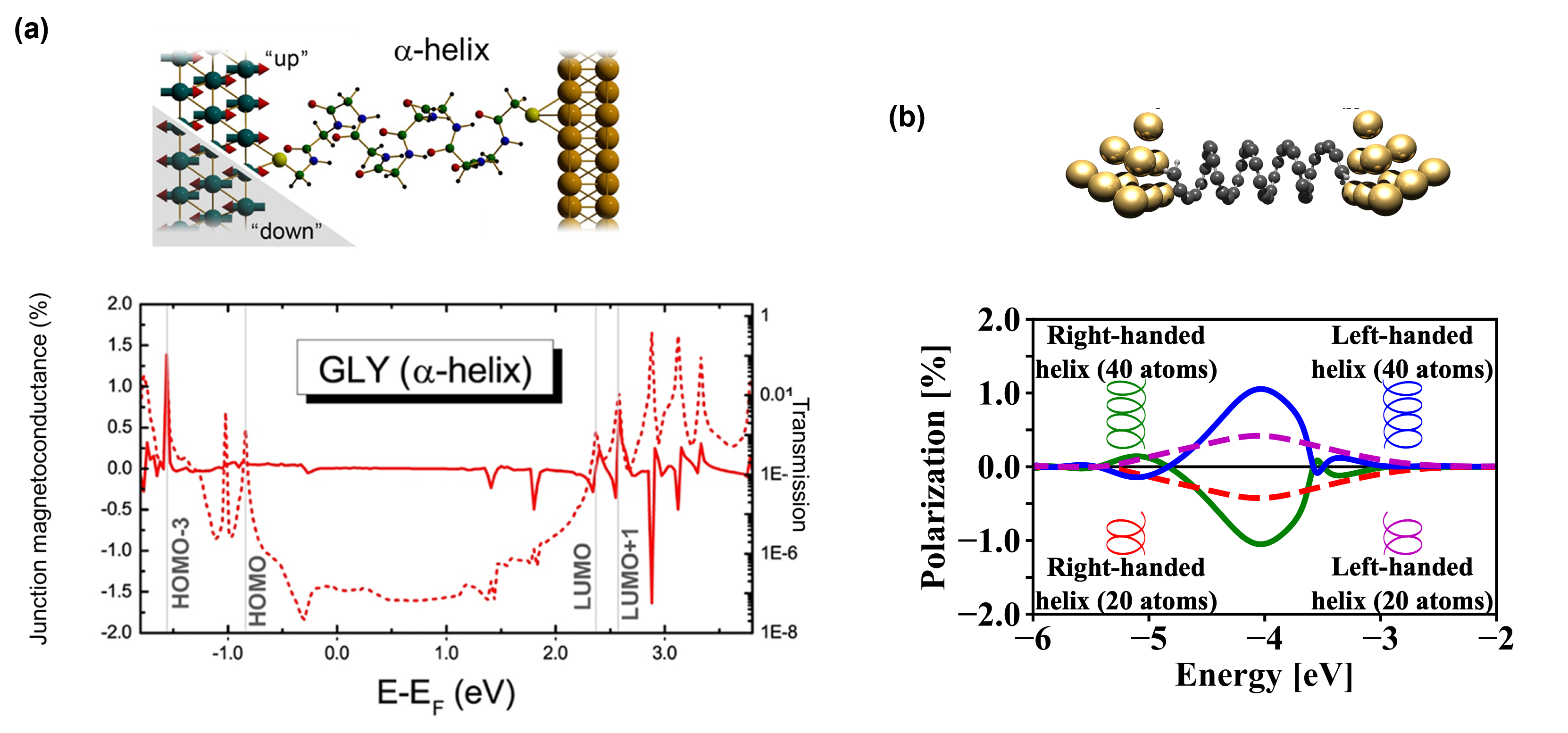} 
  \caption{(a) For a realistic peptide helix, a DFT-based Landauer approach including SO coupling yields spin polarization as rather narrow peaks far from the Fermi energy (solid line in the plot, reported as junction magnetoresistance; using PBE functionals).\cite{Maslyuk2018}  (b) For a model helix of equidistant carbon atoms (capped by two hydrogens at each end), spin polarization over a broad energy range close to the Fermi energy  is obtained, but it can be traced back to SO transfer from the gold electrodes rather than resulting from SO intrinsic to the helix (in the plot, the Fermi energy is between \nobreakdash-5~eV and -4 eV for gold; using B3LYP functionals).\cite{Zoellner2020} Note that the exchange-correlation functional PBE (plot in (a)) features 0\% Hartree-Fock exchange, while B3LYP (used on the right) has 20\%, and that the absolute values of spin polarization depend on this exchange admixture. Importantly, the polarization changes its sign when the helicity is inverted, and increases with molecular length (plot in (b)). Reprinted with permission from Refs. \citenum{Maslyuk2018} and \citenum{Zoellner2020}. Copyright 2018 and 2020 by the American Chemical Society.}
  \label{FigDFT1}
\end{figure}

\par In spite of the attractive features of CISS theories within a first-principles framework, first-principles calculations on complex helical structures and assemblies can be challenging \cite{Maslyuk2018, Zoellner2020}. Most standard implementations of DFT can treat periodic and/or finite molecular systems, with hundreds or even thousands of atoms in the simulation cell. The computational time for ground-state calculations using local or semi-local exchange correlation functionals scales with the cube of the number of atoms, while those employing Hartree-Fock or hybrid exchange scale with the fourth power. In view of these difficulties, systematic first-principles methods that treat the helical symmetry more exactly \cite{banerjee2021ab, banerjee2021, Ghosh2019a} (and therefore employ only a minimal unit cell to represent the system being studied), besides efficient methods for computing exchange interactions \cite{lin2016, Hu2017} within such a symmetry-adapted framework, are likely to emerge as powerful tools for DFT analysis of the CISS effect in complex structures.

\par The development of first-principles descriptions of molecular properties profits from quantitative benchmark experiments\cite{Vladimiro2018}. As is often the case in molecular electronics and spintronics, the lack of detailed atomistic control makes it extremely challenging to come up with such quantitative benchmark experiments for CISS. Therefore, it is critical to have access to systematic experimental studies that probe the dependence of spin polarization on molecular structures and their arrangements, as recently provided in Ref.\ \citenum{Mishra2020a}. Together with further related studies, \textit{e.g.}, on the role of local \textit{vs.}\ axial chirality or on subtle structural modifications \textit{via} chemical substituents or heteroatoms, benchmark experiments could provide a strong foundation on which to develop of a first-principles theory of CISS. 

\par
\begin{figure}[ht!]
   \includegraphics[width=3in]{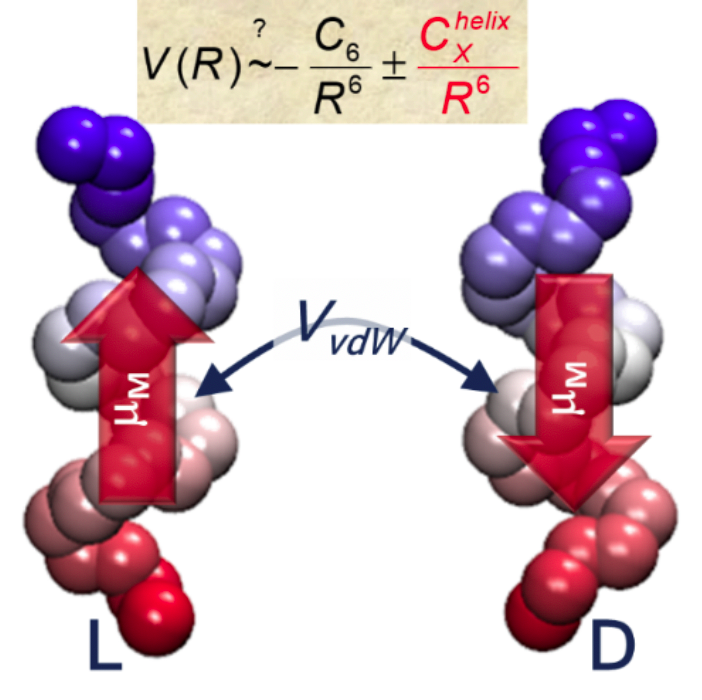} 
   \caption{In the case of chiral molecules, induction \textit{and} dispersion forces encoding electric dipole-dipole interactions require additional modifications to account for exchange-mediated interactions related to the CISS effect.}
   \label{FigDFT2}
 \end{figure}

 Aside from their relevance to spin transport physics or Heisenberg-type spin-spin interactions with magnetic substrates, helicity-dependent effects may lead to reconsideration of standard descriptions of induction and dispersion forces between chiral molecules. In analogy to CD, both electric dipole and magnetic dipole contributions are required to treat these forces, allowing for chirality-sensitive forces \cite{Craig1971, Jenkins1994} with potentially significant consequences in \textit{e.g.}, the study of intermolecular interactions in biology. These considerations are summarized schematically in \textbf{Fig.\ \ref{FigDFT2}}.

\par Overall, first-principles descriptions of CISS have made tremendous progress, but still underestimate the effect by several orders of magnitude. They might profit from developments in the areas of analytic and tight-binding theories.

\subsection*{Modeling CISS with Simple Hamiltonians: Tight-Binding and Analytical Approaches}

\par Efforts to explain the CISS effect observed in single peptide molecules with well-defined chirality (\textsc{l}- and \textsc{d}-peptides) were recently put forward, as shown in Fig.\ \ref{FigDFT1}. The transport model was based on the Landauer regime and used a Green's function technique. In the presence of spin polarization, the conductance is spin dependent and the transmission contains information associated with the molecular chirality, helicity, and the spin propagation direction. This theoretical model explained four possible scenarios of the observed current asymmetries in single chiral molecular junctions sandwiched between a polarized Ni tip and Au electrodes \cite{Aragones2017}. These four different scenarios with different conductance show that spin rectification applications close to the zero-bias limit might be possible. 


\par Using the Dirac Hamiltonian for a free electron, an intriguing prediction was made in the non-relativistic limit. The effect arises from a magnetic field, and is orders of magnitude larger than the quantum mechanical Zeeman shift.\cite{Kurian2018} This could be the source of the experimental gap in the computed spin polarization found in current CISS theories. In the low-mass approximation for free electrons (where chirality coincides with helicity), they derive a symmetry for such a system that is evocative of orbital angular momentum conversion demonstrated in vortex beams, but employing the expectation values for the energy and chirality shifts. This so-called chirality-energy conversion appears to arise from fundamental magnetic symmetries of free electrons under the influence of static fields, and such mutually correlated changes in energy and chirality can be directly measured in nano-, meso-, and macro-scale systems. A simple example of this sensitive dependence has been demonstrated in the chirality of nascent crystals and low-energy fluctuations introduced by perturbing the crystallization solution \cite{Singh1990}.


\par Understanding the physical basis of CISS \cite{Naaman2015} continues to be a challenge. Nevertheless, reduced models for the spin-polarization mechanism have helped explore distinct physical scenarios. These models involve empirical potentials or tight-binding models and can add complexity in a systematic manner.

Theoretical approaches to CISS began with attempts to explain experiments of chiral molecules in the gas phase \cite{Farago1981, Blum1989}. The theory recognized spin polarization as a single--molecule effect, where the spin-active coupling is the SO coupling between the scattered electron and the nuclear potential \cite{Blum1989}. The theory was based on symmetry considerations and geometry of the target-molecule system and agreed well with experiments,\cite{Mayer1995} but the effect was very small: the polarization asymmetry was only ${\sim}10^{-4}$.  It was then a tantalizing surprise when Ray \textit{et al}.\ \cite{Ray1999} reported a much larger effect in chiral self-assembled monolayers (SAMs) of amino acids. Minimal models using the Born series were proposed to explain this much larger effect\cite{Yeganeh2009}. A model of double scattering of single molecules, hypothesizing a SO coupling arising from C, N, and O atoms, produced chirality-dependent spin polarizations of a few percent. One of the most striking  predictions of the theory was the existence of energy windows for optimal action of the SO coupling,\cite{Vladimiro2015} which was later corroborated by Rosenberg \textit{et al}.\ \cite{Vladimiro2013}. More sensitive experiments on DNA SAMs\cite{Goehler2011} showed extraordinary electron polarizations, so the theory underestimating the CISS effect by a factor of 10. No further improvements of the theory in this regime have been realized. 

Further experimental progress accessed single molecule measurements,\cite{Xie2011} and simple tight-binding minimal models were proposed,\cite{Cuniberti2012, Guo2012} assuming quantum coherence and large SO coupling as an adjustable parameter to fit the large polarizations reported experimentally. A further step included the geometry of the orbitals and the atomic source for the SO coupling,\cite{Medina2016} using an analytical Slater-Koster approach that identified the transport SO coupling as a first-order effect in the helical geometry. Such spin coupling goes to zero in non-chiral geometries and obeys time-reversal symmetry with eigenfunctions coming in Kramer pairs.\cite{Bardarson2008} An important ingredient of the minimal model was to include the problem of the electron-bearing orbital filling in determining the energy dispersion of the model. The mechanism for spin polarization in the presence of bias is due to time-reversal symmetry being broken and a spin direction being preferred, corresponding to one of the helicity eigenstates of the Hamiltonian.\cite{Medina2016,Vladimiro2018,Geyer2020}

\par 
A minimal model for DNA, with details of the relation between geometry and transport eigenstates, recently led to a proposal that mechanical deformations may be used to assess the orbital involvement in spin transport\cite{Salazar2018}. 

\par The SO coupling is apparently the spin-active ingredient\cite{Vladimiro2015} for the molecular length-dependent of spin polarization processes. Minimal models place the strength of this coupling and its atomic origins in the range of 1-10 meV\cite{Medina2016} and the effect may be modulated by orbital overlap effects\cite{Medina2016, Gianaurelio2019} and by hydrogen bonding (as an additional source of the electric field scaling-up the effective SO coupling, particularly for biological molecules, \textit{e.g.}, DNA and polypeptides)\cite{Medina2019}.  

\begin{figure}[!ht]
  \includegraphics[width=5in]{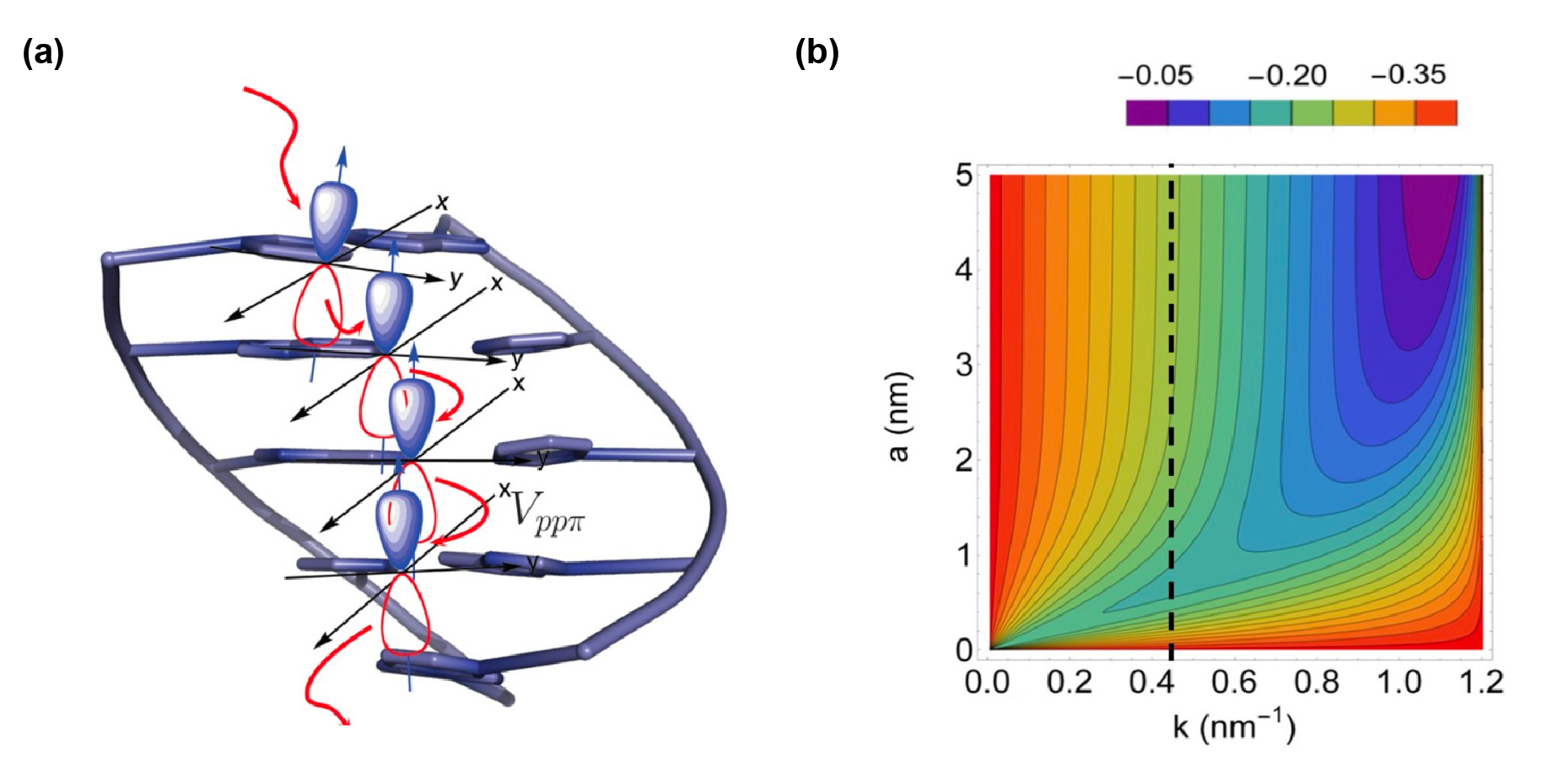} 
  \caption{\textbf{SO interaction and spin selectivity for tunneling electron transfer in DNA}. \textbf{(a)}  Schematic of DNA molecule with orbitals for electron transport. The $p_z$ orbitals are  perpendicular  to  the base planes and coupled by $V_{pp\pi}$ Slater-Koster matrix elements. \textbf{(b)} Plot of spin asymmetry $P_z$ as a function of scattering barrier length $a$ and input momentum $k$. Adapted with permission from Ref.~\citenum{Varela2020}. Copyright 2020 by the American Physical Society.}
  \label{FigTheory}
\end{figure}

Future models will need to include the two main mechanisms involved in charge transport in chiral biological molecules and polymers: tunneling and hopping\cite{Nitzan2001,Gianaurelio2010}. Which mechanism dominates transport depends on the coherence of the transport and Peierls-like instabilities\cite{Majid2020}. A simple model for nearly filled orbitals proposed a mechanism for CISS based on spin-selective transmission under a barrier\cite{Varela2020}, as shown in \textbf{Fig.\ \ref{FigTheory}}. The proposed mechanism is based on the torque-coupled interplay between electron spin and molecular angular momentum; both tunneling and hopping could be strongly influenced by spin polarization modifying the effective length of electron transfer. A recent response \cite{wohlman2020}, however, suggests that the proposed model does not lead to CISS or any net spin polarization, and further theoretical considerations are still needed to unambiguously describe CISS.

\par All models proposed for CISS focus on some extent of quantum coherence, although non-unitary proposals have been considered.\cite{Matityahu2016} CISS occurs at room temperature; how do coherent quantum mechanical effects survive? Generic voltage--probe leakage is a minimal model to examine decoherence effects.  In fact, such probes have been shown to be a route to spin polarization as breaking time reversal symmetry would indicate. A more explicit model for decoherence would include the electron--phonon coupling as a major decoherence mechanism\cite{Galperin2007,Matityahu2016, Peralta20} and describe how electronic and spin degrees of freedom couple to the phonon bath in the tunneling process. Analysis finds a partial electron--phonon decoupling to first order in the longitudinal modes \cite{Suzuura2002}, while the coupling persists for optical modes in DNA. This mechanism could also operate in oligopeptides. All of the theoretical and computational tools associated with this approach are prepared to assess the influence of spin--phonon coupling in electron transfer and transport in chiral molecules.

\subsection*{Electron Transfer and Helicity Calculations}

\begin{figure}[!ht]
  \includegraphics[width=3in]{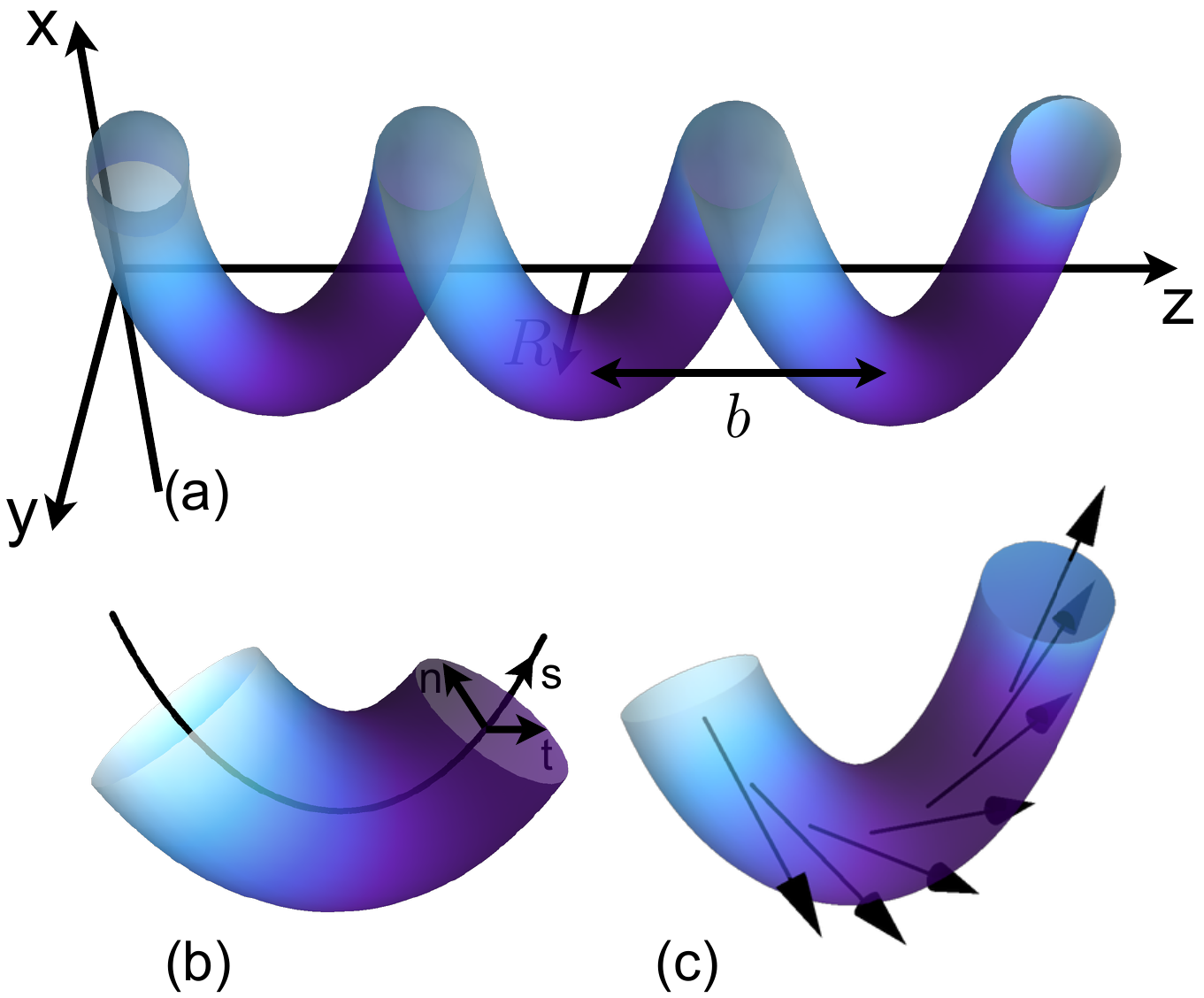} 
  \caption{\textbf{The helical tube.} (a) Electrons are confined to a helical tube of radius $R$ and pitch $b$. (b) $s$ is the position alone the helix tube and vectors $n$ and $t$ span the plane perpendicular to $s$. (c) A term in the Hamiltonian acts as an effective Zeeman field rotating as a function of the position along the helix. Adapted with permission from Ref.~\citenum{namaanEtal}. Copyright 2019 American Chemical Society.}
  \label{Fig:helix}
\end{figure}

\begin{figure}[!ht]
 \begin{flushright}\begin{minipage}{1\textwidth} \centering
       \includegraphics[width=0.95\textwidth]{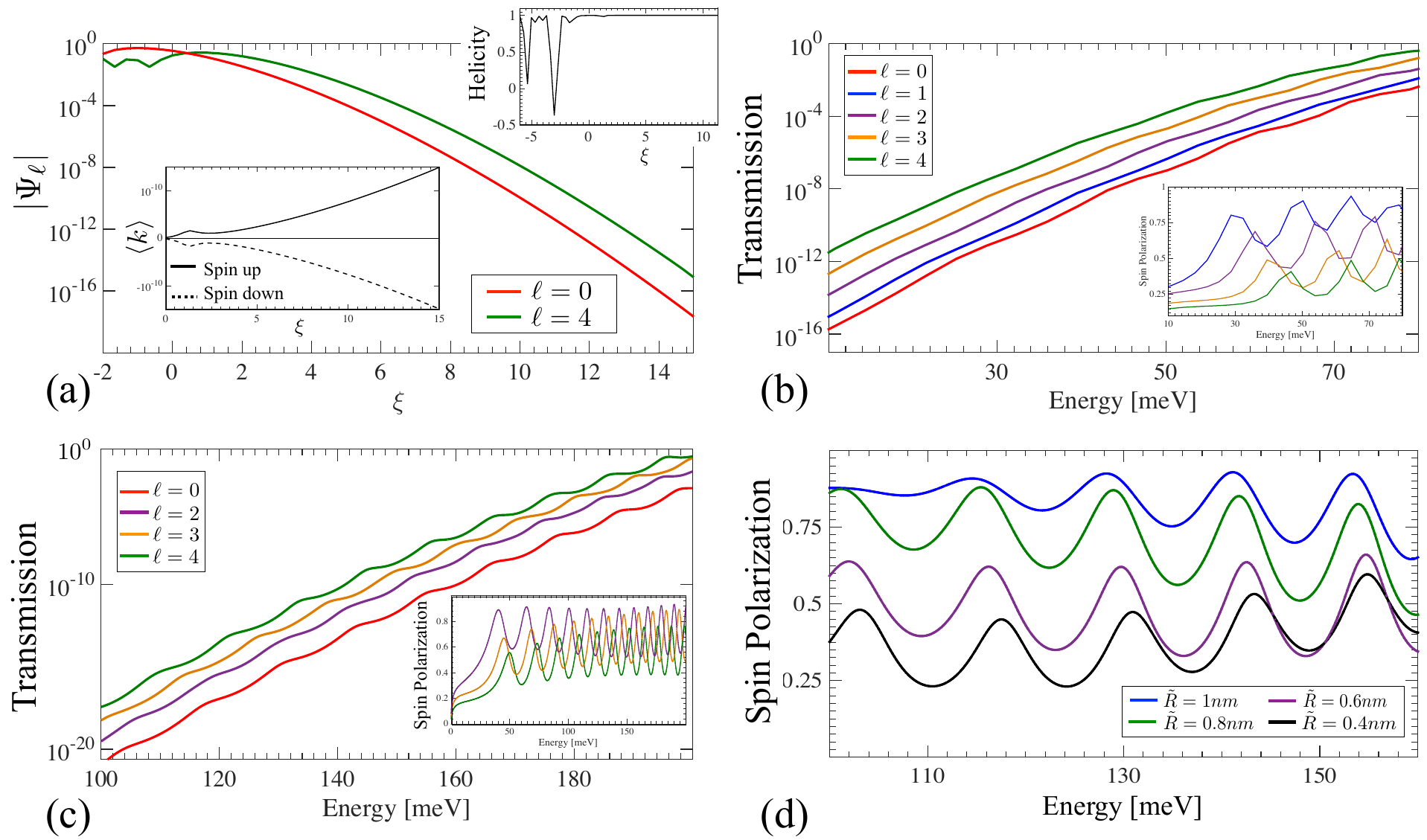}
                \caption[]{\small \textbf{Transmission through an helix-shaped molecule in the presence of a dipole field.} (a) In the presence of SO coupling, the amplitude of the (exact) electronic wave function in its tail (parametrized by $\xi\gg1$) grows as a function of the angular momentum.  Moreover, the spin is aligned along the momentum direction (see inset), and as a consequence the state has a well defined helicity. (b) The increased amplitude deep inside the molecule gives rise to an enhanced transmission probability that grows with the angular momentum $\ell$.  The scattering matrix\cite{Entin} is derived for the exact wave function. 
                This panel shows that the enhanced transmission for $\ell\neq0$ is accompanied by a spin polarization (inset). 
                (c) Similar results were obtained using a tight binding calculation for a molecule with the same parameter but somewhat different length. 
                (d) Deforming the molecule to have a larger pitch or radius helps  spin polarization. 
                Adapted with permission from Ref.~\citenum{namaanEtal}. Copyright 2020 American Chemical Society.}
    \label{Fig:amp_tra}
\end{minipage}\end{flushright}
\end{figure}

Recent theoretical attempts to explain the CISS effect and large tunneling-based transfer through chiral molecules have relied on the large SO coupling, which is uncommon in organic materials. However, Naaman \textit{et al.}\cite{namaanEtal} have shown that the chiral geometry of the molecules induces correlations between the electron spins and their flow direction and this accounts for the spin selectivity of these molecules, see Fig.\ \ref{Fig:helix}. Moreover, by adding an overall dipole potential along the molecule, it is possible to obtain enhanced electron transport, as depicted in Fig.\ \ref{Fig:amp_tra}. 

%

After finding the electronic states inside the molecule, one can calculate the local helicity by calculating the local velocity of each spin. The results indicate that, for a right-handed helix, the local helicity of the spin up (down) is always positive (negative) in the tail of the wave function. Therefore, the spin selectivity effect takes place. 


When there is no dipole coupling, the energy window where strong spin-dependent transport is observed is determined by a partial energy gap $\sim$ 1 meV. When a dipole potential is introduced, this energy window is instead determined by the dipole energy $\sim$ 0.1 meV  and the transmission probability of polarized electrons is enhanced (see Fig.~\ref{Fig:amp_tra}).  This is in agreement with experimental observations that the total transmission decreases with increasing molecule length while the spin polarization increases. 

\section*{Engineered Chiral Materials} \label{materials}

\subsection*{Influence of the Substrate SO Interaction on Experimental Spin Polarization and CISS}

\indent \par Early CISS experiments were performed predominantly on gold substrates because the helical biomolecules studied could be readily thiolated on one end to form a strong bond with the substrate. Heavy metal substrates such as gold emit electrons with a preferential spin orientation when excited by circularly polarized light. The spin orientation of the emitted electrons is linked to the helicity of the light. In noble metals, the \textit{d}-electron SO coupling constants are measured to be 0.1, 0.25, and 0.72 eV for Cu, Ag, and Au, respectively.\cite{Woehlecke1983}  On single--crystal Au(111), circularly polarized ultraviolet radiation just above the work function excites the $\Lambda^1_6 \leftarrow \Lambda^3_4\Lambda^3_5$ electronic transition, from the spin-polarized occupied band into an unoccupied plane--wave final band $(\Lambda^1_6)$  near the L point of the Brillouin zone. For Cu and Ag,  in addition to the lower SO coupling, the initial state is a $\Lambda^3_6$ state that produces only weak spin polarizations.\cite{Woehlecke1983} Such an excitation with circularly polarized light then yields longitudinally polarized electrons with respect to the quantization axis (\textit{i.e.}, the $k$ vector of the exciting radiation). It is thus important to excite the system at normal incidence and also to extract electrons normal to the surface. For Au(111), spin polarization values up to $P=30\%$ are obtained just above the vacuum level.\cite{Woehlecke1983, Goehler2011} Due to the electronic structure of polycrystalline gold, the spin direction is reversed compared to Au(111), and the magnitude is also significantly smaller. Nevertheless, this substrate behavior suggested that the strong SO interaction in heavy metals extends to the helical adsorbates that consists of light atoms, including C, N, O, and H.\cite{Gersten2013} 

\par
\begin{figure}[ht!]
  \includegraphics[width=6in]{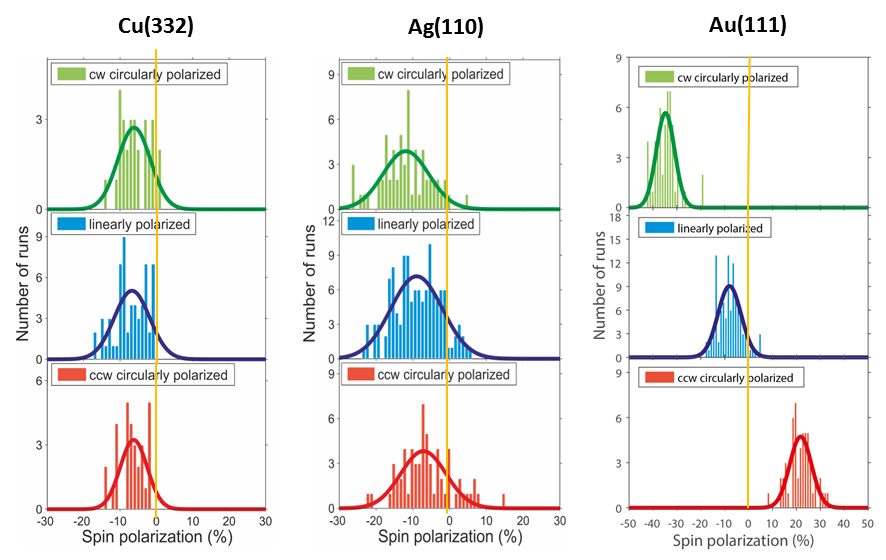} 
  \caption{\textbf{Spin polarization of photoelectrons from Cu, Ag, and Au substrates transmitted through a monolayer of M hepta-helicene.} Green, blue, and red histograms (from top to bottom) represent excitation by clockwise (cw) circularly, linearly, and counterclockwise (ccw) circularly polarized light at $\lambda=213$~nm, and thus emitting electrons slightly above the vacuum level of the systems. Adapted from Ref. \citenum{Kettner2018}.}
  \label{FigHelicenes}
\end{figure}

 For these three noble metals, Au, Cu and Ag, systematic CISS experiments were performed for adsorbed monolayers of enantiopure hepta-helicene.\cite{Kettner2018} For linearly polarized exciting light that produces unpolarized photoelectrons in the substrate, M-helicene yields a spin polarization of $P=-6.7\%$ on Cu(332), $P=-9.0\%$ on Ag(110), and $P=-8.0\%$ for Au(111) (see \textbf{Fig.~\ref{FigHelicenes}} center (blue) histograms). Circularly polarized excitation that already produces polarized photoelectrons in Au and, to a limited extent, also in Ag substrates, generates additional spin polarization. This effect is particularly noticeable for Au(111). There, clockwise (cw) circularly polarized light (upper green histograms) yields a total spin polarization of $P=-35\%$ for M-[7]-helicene, while counterclockwise (ccw) polarized light (lower red histograms) produces $P=-22\%$. For P-helicene, the sign of the spin polarization switches, and the action of cw- and ccw-polarized light on the total spin polarization is also reversed.\cite{Kettner2018}

\par It was shown that the CISS effect occurrs for bacteriorhodopsin adsorbed on aluminum oxide\cite{Mishra2013} and for DNA bound to Si(111);\cite{Kettner2016} both systems lack significant SO coupling. The influence of the helical organic molecules on CISS is thus independent of the substrate. However, an initial spin orientation may contribute to reaching high spin polarization. It is thus promising that neither heavy substrate elements nor magnetic substrates are required to generate spin-polarized electrons from helical organic molecules. These results lift possible restrictions in designing spintronic elements, as well as motivates the application of CISS in (electro-)chemistry\cite{Ghosh2019} and to biological systems.\cite{Mishra2013}


\subsection*{Heavy-Element Insertion}

\indent \par Early studies of spin-dependent attenuation of electron beams through vapor-phase chiral molecules by Mayer and Kessler found that the presence of a heavy element, such as ytterbium, bound to a chiral molecule could significantly enhance the transmission asymmetry \cite{Mayer1995}. These findings agree qualitatively with earlier theoretical models, indicating that the presence of a heavy atom in a chiral molecular environment should enhance the spin- and chirality-dependent asymmetry in electron-molecule interactions, likely due to increased SO coupling effects \cite{Blum1989}. These results were further expanded using chiral bromocamphor derivatives \cite{Mayer1996}. Subsequent experiments found that the degree of spin-polarized electron transmission asymmetry could be modified for nearly identical molecules simply by substituting the coordinating species. Spin-selective electron transmission asymmetry through vapors of chiral camphor derivatives was observed to increase roughly in proportion to the atomic number of the coordinated atom, as Pr (Z = 59) $<$ Eu (Z = 63) $\sim$ Er (Z = 68) $<$ Yb (Z = 70) \cite{Nolting1997}. Interestingly, molecules with multiple heavy atom inclusions, such as dibromocamphor, did not exhibit higher asymmetries than their singly brominated counterparts. These trends are also found in recent studies carried out on single-stranded DNA SAMs designed to form DNA hairpins that coordinate Hg$^{2+}$ ions at thymine--thymine mismatch sites \cite{Stemer2020}. 

\begin{figure}[ht!]
    \includegraphics[width=5.5in]{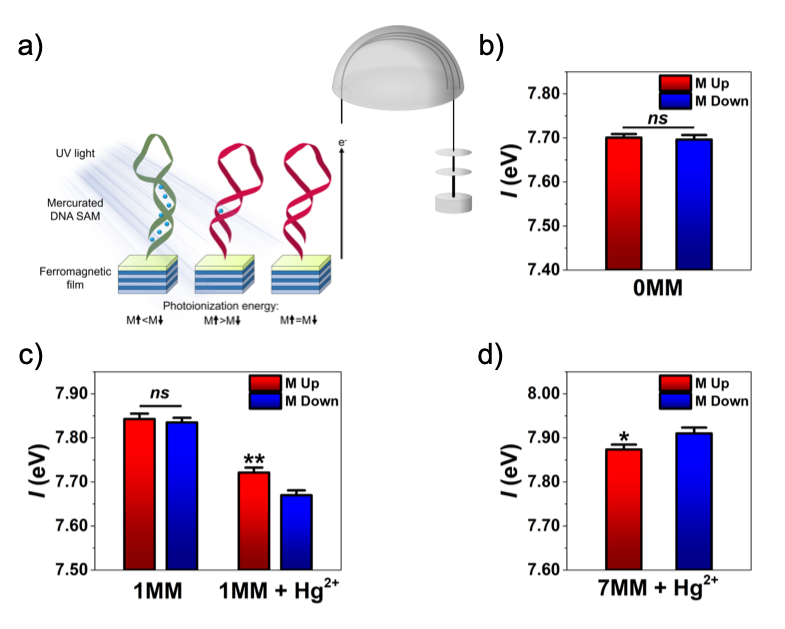}
    \caption{\textbf{(a)} Schematic depicting spin-dependent photoelectron scattering through self-assembled monolayers of DNA hairpins on ferromagnetic films, characterized by ultraviolet photoelectron spectroscopy. Spin-dependent ionization cross sections result in differential charging, physically manifested as substrate magnetization-dependent photoionization energies of the chiral organic films. \textbf{(b,c)} Spin-selective effects were only observed in short ($\sim$1 helical turn) DNA hairpins that contained mercury bound at thymine--thymine mismatches due to enhanced molecular SO coupling. \textbf{(d)} Spin selectivity was reversed in DNA hairpins containing 7 mismatches and stoichiometric amounts of mercury ions, which was shown to invert the chirality of the helical hairpins. Reproduced with permission from Ref.~\citenum{Stemer2020}. Copyright 2020 by the American Chemical Society.}
    \label{FigSpinOrbitHg}
\end{figure}

\par Employing ultraviolet photoelectron spectroscopy to characterize magnetization-depend\-ent ionization energies of DNA SAMs formed on ferromagnetic substrates, Stemer \textit{et al}.\ (see \textbf{Fig.\ \ref{FigSpinOrbitHg}}) found that incorporating a single equivalent of Hg$^{2+}$ in DNA hairpins with only a single helical turn is sufficient for the manifestation of spin-dependent effects at room temperature. No magnetization-dependent effects are apparent in the samples composed of identical DNA without Hg$^{2+}$. At high metal loading, the tertiary structure of the DNA hairpins was found to invert. This inversion was accompanied by a corresponding reversal in the preferred magnetization orientation for photoionization. Analogous to earlier experiments by Kessler and Mayer\cite{Mayer1995}, increased incorporation of heavy elements did not further increase spin-dependent interaction asymmetries, indicating that multi\-ple heavy inclusions may induce compensating rather than amplifying effects. Recent studies on peptides incoporating paramagnetic inclusions reach similar conclusions with respect to increase of spin polarization effects.\cite{Cavanillas} These studies highlight the tunability of chiral molecular systems \emph{via} the incorporation of heavy species, a powerful tool in engineering highly spin asymmetric systems for spintronics and quantum computing applications.

\subsection*{Functionalization-Induced Chirality in 2D Materials}

\begin{figure}[htbp]
\centering
\includegraphics[width=6in]{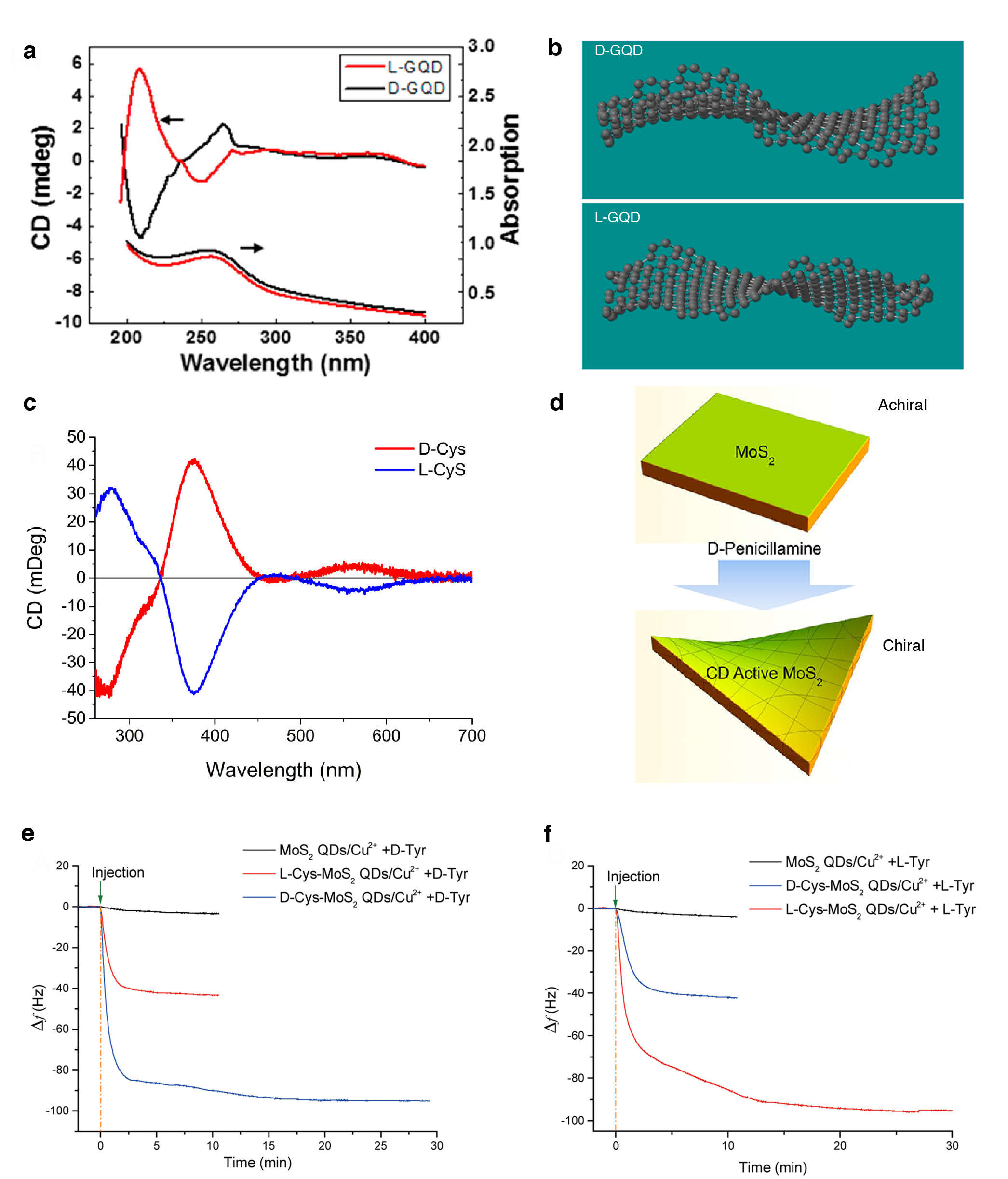}
\caption{\label{fig1}
(a) Circular dichroism and optical absorption spectra for \textsc{l}-GQD (red) and \textsc{d}-GQD (black)
dispersions. (b) Schematic of the structures of D-GQD and \textsc{l}-GQD on a 3 nm QD. The rotation direction 
of the helices is opposite to the handedness of the edge-ligands. 
(c) Circular dichroism spectra of \textsc{d}/\textsc{l}-cysteine functionalized MoS$_2$ after exfoliation. 
Similar results for \textsc{d}/\textsc{l}-penicillamine were obtained (not shown here).
(d) Diagram of the deformations generated by molecular 
functionalization of \textsc{d}-penicillamine inducing chirality on MoS$_2$ layers.  
(e--f) Frequency shifts of the \textsc{d}- and 
\textsc{l}-Cys–MoS$_2$ QDs/Cu$^{2+}$ sensors exposed to (e) 5 mM \textsc{d}-Tyr and 
(f) 5 mM \textsc{l}-Tyr solution. Adapted with permission from Refs. \citenum{Suzuki:2016aa, Purcell-Milton:2018aa, Zhang:2018aa}. 
Copyright 2016 and 2018 by the American Chemical Society. 
}
\end{figure}

Chirality in 2D materials has spanned over different compounds from graphene QDs\cite{Suzuki:2016aa} and nanodisks\cite{Kong:2017aa}, 
nanoplatelets of CdSe\cite{C8NR10506E}, MoS$_2$ layers\cite{Purcell-Milton:2018aa}, 
up to colloidal semiconductor quantum wells of different types\cite{doi:10.1002/adfm.201802012} and 
transition-metal dichalcogenide semimetals\cite{Xu:2020aa}. 
One of the earliest systems to show chiral properties in strictly 
two dimensions was graphene flakes or QDs (GQD) covalently 
modified by \textsc{l}- or \textsc{d}-cysteine moieties\cite{Suzuki:2016aa}. 
By using edge modifications \emph{via} aqueous dispersions, 
Suzuki \textit{et al.} demonstrated that chiral amino acids can 
induce excitation bands with specific features in the circular dichroism spectra
at 210$-$220 nm and 250$-$265 nm (Fig.\ \ref{fig1}{\bf a}). Comparing the optical 
responses for \textsc{l}/\textsc{d}-cysteine bound to GQDs (\textsc{l}-GQD and \textsc{d}-GQD) a clear asymmetry is noticed at the 
new bands which can be correlated to structural deformations of the graphene layer itself. Once 
the ligands were linked to the edges an increase in the buckling deformation of the GQD 
was observed. That is, depending on what amino acids is present at the edges, the GQDs have 
a right- or left-handed twists (Fig.\ \ref{fig1}{\bf b}). 
The buckling is due to noncovalent intermolecular interactions of the amino acids 
with each other and also with other parts of the edge. Indeed, it is well known that 
deformations at chiral regions of large molecules tend to propagate 
the chirality through other regions or host environments such as 
proteins, polymers, liquid crystals\cite{C4CS00531G,Liu:2015aa}. 
With L/D-GQDs showing chirality properties, one can explore 
different pathways for biological recognition in functional cells, 
\textit{e.g.}, neurons\cite{doi:10.1002/adfm.201805512,TOSIC201995,D0RA00799D}, 
bone marrow cells\cite{doi:10.1002/biot.201800249,Li:2019aa,Fasbender:2019aa,doi:10.1080/21691401.2019.1576706}, or 
immune cells\cite{Leeeaaz2630,TOMIC201713}. 
In particular, L/D-GQDs show biocompatibility 
with human liver hepatocellular carcinoma cells
where different toxicity was observed\cite{Suzuki:2016aa}. 
Even though it is not totally clear why such difference 
emerged, a few factors in terms of atomic chirality of the ligands and 
the twist of the graphene sheets may play a role. 

Similar approach using liquid exfoliation  
with chiral molecules, \textit{e.g.}, \textsc{l}-cysteine and \textsc{d}-penicillamine, 
resulted in chiral thin-layers of MoS$_2$\cite{Purcell-Milton:2018aa}. 
Purcell-Milton \textit{et al.} reported that after ligand functionalization  
MoS$_2$ shows strong circular dichroism signals (Figure \ref{fig1}{\bf c}). 
The optical spectra display almost a mirror image relative to the wavelength considered 
with the different molecules bound to the surface. The characteristic changes of the circular  
dichroism varying optical orientation between positive and negative signs 
as function of the wavelength is due to the Cotton effect. Moreover, other amino acids such as 
glutamic acid, alanine, and methionine, were 
also tested by Purcell-Milton \textit{et al.} but none of them 
have resulted in chirality of the MoS$_2$ layers. This suggests that 
a ligand coordination as well as binding through at least two functional 
groups should take place at the surface\cite{B808054B}. 
One of the implications to have a sizeable 
interactions between \textsc{l}-cysteine and \textsc{d}-penicillamine and MoS$_2$ is the structural 
deformation of the nanosheets with the functionalization (Figure \ref{fig1}{\bf d}). 
That is, an achiral MoS$_2$ layer can be transformed into a chiral sheet. This argument 
was used to model the optical spectra of both molecules on MoS$_2$ resulting in 
close agreement with the experimental results. 
 
The utilization of liquid exfoliation using chiral molecules provides 
a large scale production of transition metal 
dichalcogenides with chiral features. 
In principle such strategy can be extended to many layered materials
(\textit{e.g.}, MoSe$_2$, WS$_2$, and WSe$_2$) with applications that can extend those for 
sensing or optical devices. For instance, in enantioselective catalysis 
and peroxidase activity\cite{Zhang:2018aa}. Zhang \textit{et al.} 
demonstrated that MoS$_2$ layers functionalized with \textsc{d}-, \textsc{l}-cysteine (\textsc{d}-Cys-MoS$_2$, \textsc{l}-Cys-MoS$_2$) 
show enantioselective peroxidase activity on chiral substrates 
composed by \textsc{d}- and \textsc{l}-tyrosinol (Tyr) enantiomers (Figure \ref{fig1}{e-f}).  
With the assistance of copper ions (Cu$^{2+}$), both \textsc{d}-Cys-MoS$_2$ and 
\textsc{l}-Cys-MoS$_2$ systems display high oxidation of \textsc{d}-Tyr (Fig.\ \ref{fig1}{\bf e})
and \textsc{l}-Tyr (Fig.\ \ref{fig1}{\bf f}) beyond the unmodified MoS$_2$ QDs. 
Indeed, no peroxidase activity is observed for times within 
10 minutes for pristine MoS$_2$. This suggests that modified 2D materials 
can function as nanozymes with enantioselectivity.

\subsection*{\textcolor{black}{Spin Selection of Circularly Polarized Light in Chiral Molecule--Nanoparticle Hybrids Mediated by Resonant Plasmonic Fields}}

The CISS effect describes the spin filtering and spin polarization of electrons in transport, transfer and bond polarization processes. It is mediated by chiral molecules and chiral interfaces, which break space inversion symmetry, provided time-reversal symmetry is broken. Interestingly, similar effects occur in other physical phenomena such as in the propagation of light in the presence of chiral matter. Although their proper understanding came later, these effects have been known since the discovery of optical rotation by Arago and Biot in the early 19th century and the first observations of circular dichroism by Wilhelm Karl von Haidinger. By representing linearly polarized light as a coherent superposition of left-handed (spin +1) and right-handed photons (spin -1), the optical rotation can be rationalized as a dephasing of left-handed photons with respect to the right-handed photons due to their different indices of refraction when traversing chiral materials. Similarly, the Circular Dichroism, CD, can be understood as the preferential absorption of photons of one particular spin in the interaction with chiral matter\cite{Barron2004-wr}.

The spin selectivity in CISS is measured by the spin polarization, $S$, defined as $S=\frac{I_+-I_-}{I_++I_-}$, where $I_+$ and $I_-$ are the signal intensities of the transmitted electrons with the spin parallel and antiparallel to their linear momentum vector, respectively. Similarly, since many light-matter interaction processes are governed by the differential absorption intensity, it is customary to refer to the Khun’s dissymmetry parameter $g=\frac{A_+-A_-}{A_++A_-}$, where $A^+$ and $A^-$are the absorption intensities of the chiral material when illuminated by left and right polarized incident fields respectively. The absorption intensity of left ($+$) and right ($-$) handed circularly polarized fields by a chiral molecule can be expressed as\cite{Tang2010-mn,Poulikakos2019-um}:

\begin{equation} \label{A}
A_{\pm} = \frac{\omega}{2}\left( \alpha^{''} \left|\vec{E} \right|^2 +\mu^2_0 \chi^{''} \left|\vec{H} \right|^2 \right) \mp \frac{2}{\varepsilon_0}G^{''}C,
\end{equation}

where $\vec{E}$ and $\vec{H}$ are the complex electric and magnetic fields, $\omega$ is the frequency of light, $\mu_0$ is the permeability of free space and $\alpha^{''}$, $\chi^{''}$ and $G^{''}$ are the imaginary part of the electric, magnetic and chiral polarizabilities of the chiral molecule. $C$ is the so-called electromagnetic density of chirality, 

\begin{equation} \label{C}
C = \frac{\omega}{2c^2}Im\left( \vec{E}^* \cdot \vec{H} \right).
\end{equation}

Assuming that the $\mu^2_0 \chi^{''} \left|\vec{H} \right|^2$ is negligible due to the weak magnetic polarizability of most small molecules, combining Eqs. \ref{A} and \ref{C}, Khun’s dissymmetry parameter can be recast as:

\begin{equation} \label{g}
g = -\left( \frac{G^{''}}{\alpha^{''}} \right) \left( \frac{8C}{\omega\varepsilon_0\left|\vec{E} \right|^2} \right)
\end{equation}

Similarly, the circular dichroism spectra of molecules, defined as $CD=A_+-A_-$can also be reformulated as:

\begin{equation} \label{CD}
CD=-\frac{4}{\varepsilon_0}G^{''}C
\end{equation}

\begin{figure}[!ht]
	\centering
	\includegraphics[width=1\textwidth]{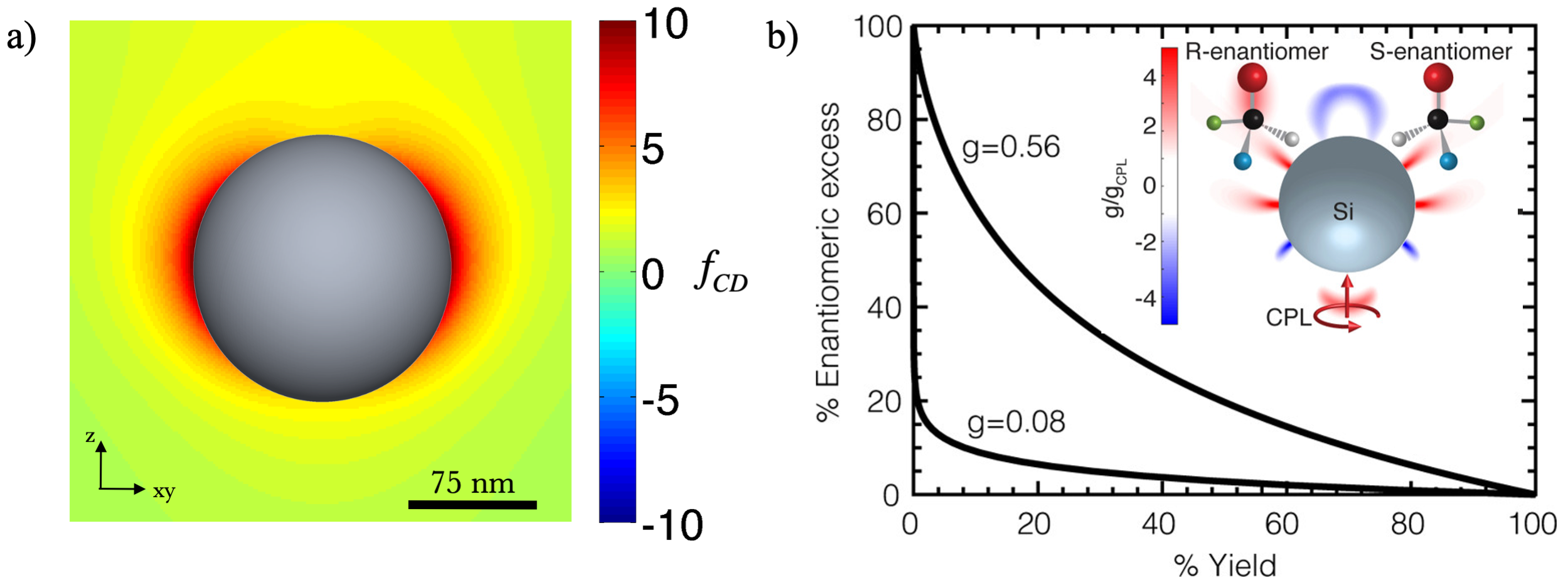}
	\caption{ a) Nanoparticles can enhance molecular CD spectroscopy. A 75 nm radius silicon nanosphere is illuminated from below by $\lambda = 625$ nm circularly polarized light. The figure depicts the CD enhancement factor at a plane crossing the center of sphere. b) When Khun’s dissymmetry factor is increased in the vicinity of a nanoparticle, the enantiomeric excess achievable in a photochemical reaction  increases dramatically for a given extent of reaction. (Inset) Spatial distribution of dissymmetry factor enhancement for a 536 nm Si sphere at $\lambda = 1391.82$ nm. In the presence of a Si nanosphere, in a region where $g\approx 7$, a $20\%$ enantiomeric excess can be reached with a yield of $50\%$ compared to the $1\%$ achievable in the absence of the nanoparticle. Panel (a) adapted from Ref. \citenum{Garcia-Etxarri2013-lx} with permission, copyright 2013 American Physical Society. Panel (b) adapted from Ref. \citenum{Ho2017-kq} with permission, copyright 2017 American Chemical Society.
	}
	\label{SECD}
\end{figure}

It is readily observed that the first term in both equations depends exclusively on the intrinsic polarizabilities of the molecules under measurement. On the contrary, the second term corresponds to a purely electromagnetic quantity, which can be engineered through optical means. This conclusion, which was first theoretically observed\cite{Tang2010-mn} and later experimentally proved\cite{Tang2011-yu} by Y. Tang and A. Cohen, triggered a myriad of studies on field enhanced chiral-light matter interactions. In particular, in the recent past, plasmonic\cite{Maier2007-xq} and all-dielectric nanoparticles\cite{Garcia-Etxarri2011-ln,Kuznetsov2016-da} have proven to be able to augment the efficiency of many chiral-light matter interactions. For instance, CD spectroscopy of chiral molecules nowadays can be enhanced by means of individual metallic\cite{Nesterov2016-nk} (plasmonic) and dielectric nanoparticles\cite{Garcia-Etxarri2013-lx} (Figure \ref{SECD}a). Moreover, it has been recently proven that arrays of high-refractive index nanoparticles\cite{Solomon2019-oi,Hu2020-mj,Lasa-Alonso2020-go} can provide one order of magnitude higher CD enhancement factors. Furthermore, the (usually very inefficient) asymmetric photoseparation of chiral enantiomers\cite{Flores1977-gt,Modica2014-oa}, can also benefit from enhanced chiral-light matter interactions. High-refractive index nanostructured materials have shown a great potential to selectively induce photolysis and ionization of particular chiral enantiomers while leaving the other ones almost unaffected\cite{Solomon2019-oi,Ho2017-kq} (Figure \ref{SECD}b). 

\begin{figure}[!ht]
	\centering
	\includegraphics[width=0.9\textwidth]{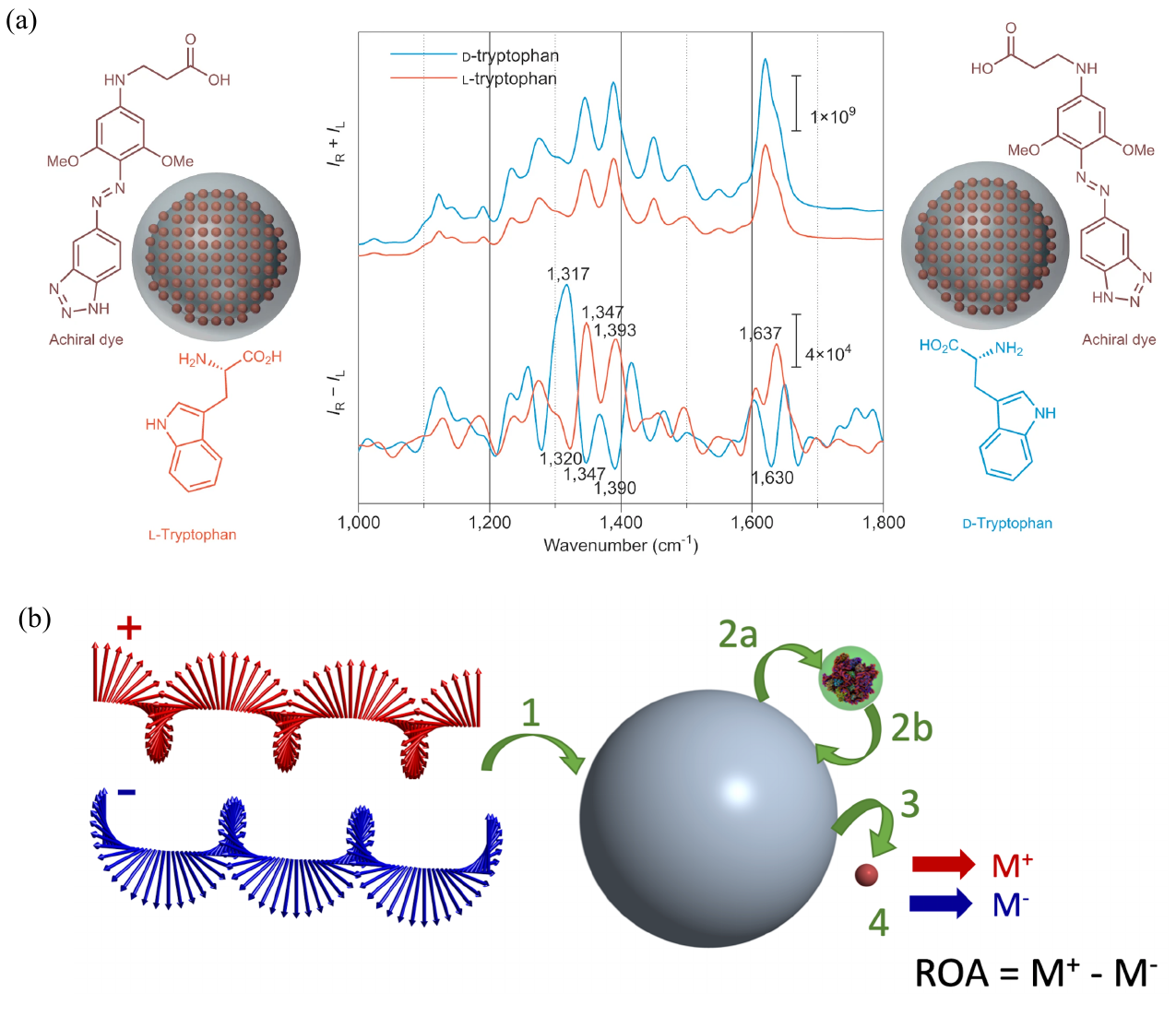}
	\caption{ (a) Schematic of the plasmonic nanoparticle--chiral molecule hybrid. The silver nanoparticle is surrounded by reporter Raman molecules, surrounded by a silica shell and has attached a chiral analyte. Middle plot shows surface enhanced resonant Raman spectra of \textsc{d}- and \textsc{l}-tryptophan bound to the benzotriazole-functionalized nanotag, which are identical as expected for the two enantiomeric systems. Surface Enhanced Raman Optical Activity (SERROA) spectra in the presence of the two chiral enantiomers. Strong chiroptical responses in several signals are observed, demonstrating the chirality transfer phenomena. (b) Schematic of the model describing the chiral molecule--nanoparticle hybrid. Step 1: circularly polarized light excites the plasmonic nanoparticle. Steps 2a and b: The plasmonically enhanced fields excite the chiral molecule, which backscatters light into the papartile, producing a self consistent chiral polarization of the plasmonic particle. Step 3: The chiral polarization of the chiral molecule--nanoparticle hybrid excites the non-chiral Raman reporter. Step 4: The Raman molecule emitis a surface enhanced resonant Raman signal, which is different for left handed ($M^+$) and right handed ($M^-$) circularly polarized incidence, creating a measurable Raman Optical Activity signal. Panel (a) Reprinted and adapted with permission from Ref. \citenum{Mujica2015} based on work in Ref. \citenum{Ostovar_pour2015-wh}, copyright 2015 Nature Publishing Group. Panel (b) adapted with permission from Ref. \citenum{Garcia-Etxarri2020-wz}, copyright 2020 American Chemical Society.}
	\label{CISSlight}
\end{figure}

Most interestingly, for the topic of this review, very recently it was discovered that electromagnetic chiral interactions could be used to transfer chirality information from a chiral molecule to an optically inactive (achiral) Raman molecule. Transfer of chirality is known to occur between molecules through chemical bonds\cite{Kumar2018-kj}. However, in 2015, Saeideh Ostovar pour et. al., showed in a seminal article that chiral information could be transferred from a chiral molecule to an achiral one in the presence of a plasmonic nanosphere\cite{Ostovar_pour2015-wh} (Figure \ref{CISSlight}a-b) irradiated by circularly polarized light of its resonant frequency. The transfer of chiral information was proven through a Raman optical activity (ROA) experiment. The chiral molecules in the experiment were Raman inactive, while the achiral molecules displayed a strong Raman response. In the absence of chiral molecules, the plasmonic nanoparticle-Raman molecule hybrid did not present any ROA signal. Nevertheless, when the non-Raman active chiral molecule was added in the system, the experiment presented a strong ROA signal\cite{Mujica2015-yc}. Since the chiral molecule and the Raman molecules were spatially well separated, this chiral information transfer could not occur through the formation of a chemical bond between the chiral and the achiral system; it had to be mediated by the plasmonically enhanced electromagnetic interactions. 

In a later article, these experiments were addressed theoretically and explained on the basis of enhanced chiral-light matter interactions\cite{Garcia-Etxarri2020-wz}. On the first step, the plasmonic nanoparticle is excited by the circularly polarized incoming beam (step 1 in Figure \ref{CISSlight}c). Then, the plasmonic resonant response of the particle produces enhanced fields in their surroundings. These enhanced near fields excite the nearby chiral molecules following Eq. 1, as depicted in step 2a in Figure \ref{CISSlight}c. Since the plasmonic nanoparticle is not chiral, overall it will not induce any additional chirality in the electromagnetic response around it. Nevertheless, the excitation of the chiral molecule in its surroundings will filter the spin of the exciting fields, absorbing more efficiently photons of one particular handedness. As a result, the back scattering of the fields from the molecule to the particle, step \ref{CISSlight}b in Figure \ref{CISSlight}c, induces a chiral self-consistent electromagnetic polarization in the plasmonic particle. Lastly, these self consistent fields will excite the non-chiral Raman molecule (Figure \ref{CISSlight}c, step 3), which will give a measurable Raman signal, denoted as $M^+$, in the far field. From the point of view of symmetry arguments, the main ingredient of this model is that inversion symmetry is broken at the interface between the chiral molecule and the plasmonic nanoparticle.

Furthermore, exciting the plasmonic-nanoparticle chiral-molecule hybrid with  a circularly polarized incoming beam of the opposite handedness, will induce a different optical response in the system due to the specific handedness of the chiral molecule. As a result, the Raman molecule, will now be excited by electromagnetic fields of a different intensity, i.e. the field has become ``chiralized'' resulting in a different Raman signal in the far field M-M+. Since the ROA signal is given as $ROA=M^+-M^-$, the non-chiral Raman molecule will now produce a measurable ROA signal due to its excitation by the radiated chirality-information carrying electromagnetic field from the plasmonic-nanoparticle chiral-molecule hybrid (step 4 in Figure \ref{CISSlight}c).

In summary, it has been shown that the hybrid system composed by the non-chiral plasmonic nanoparticle and the chiral molecule behaves distinctly with respect to the circularly left- and right-polarized incident light of the frequency of the plasmonic resonance. Indeed, since the chirality of the molecule preferentially enhances the intensity of the incident light with one spin (polarization) over the other one, the intensity of the electric part of the back radiated light’s electromagnetic field at the position of the non-chiral Raman active probe depends on the combination of the incident handedness and the chirality of the Raman inactive molecule of the hybrid system. Ultimately, this difference is sensed by the Raman active non-chiral molecule because its Raman signal’s intensity is proportional to the fourth power of the sensed electric field’s intensity.

Thus, we have seen that chirality induces spin selection of the light shining hybrid systems containing plasmonic and chiral species, and that the information about the selected spin can be transmitted by the back radiated electromagnetic field over long distances, where it can be decoded by a Raman active non-chiral molecule. This chiral-induced spin selectivity constitutes, therefore, a highly efficient method to transmit and decode information about chirality over large distances which expand to the nanoscale. The implications of this chirality transfer for the design of devices, for electron-phonon information transfer, and in the field of photo-induced spin-dependent chemistry are important.

\textcolor{black}{In the future, the use of artificial chiral nanostructures such as arrays of plasmonic and dielectric objects can be explored for their use in similar spin interactions. For example, nanoparticles, arrays of nanodisks and holes, and photonic metasurfaces have been shown to exhibit electric and magnetic resonances for enhancing chiral optical density,\cite{Solomon2020,Qin2020}} and would be excellent candidate materials for chiral spin selection. These types of nanophotonic materials have the benefit of ease of scalable fabrication, and independently tunable properties based on their materials and geometries.

\section*{Chirality in Biology} \label{bio}


\indent \par Chirality plays a fundamental role in biology, ranging across vast orders of magnitude in scale. Researchers have pondered the origins of this universal handedness from multiple vantage points, including cosmological \cite{Alexander2014}, astronomical \cite{Bailey2001}, astrophysical \cite{Cline2005,life9010029}, biomolecular \cite{Blackmond2019, Kumar2017, Kurian2018}, violation of parity symmetry \cite{Quack}, and quantum field theory \cite{Kurian2018}. The connections between spontaneous symmetry breaking, chiral effects, and life are intimate, but this nexus has not been understood.

\par
\begin{figure}[ht!]
  \includegraphics[width=3.5in]{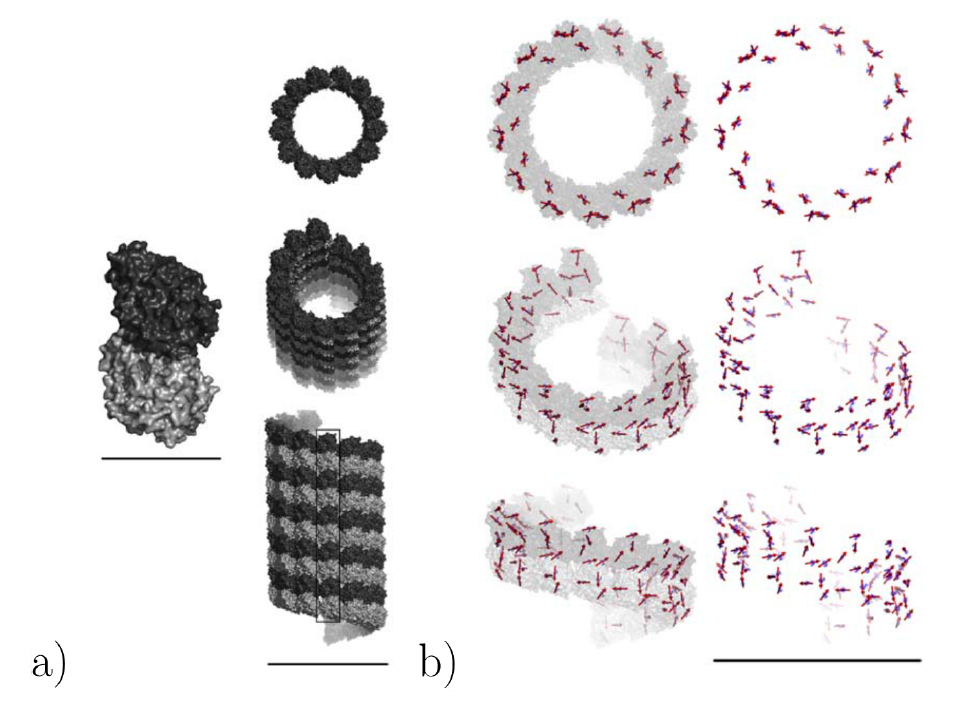} 
  \caption{(a) (left) Tubulin protein (scale bar ~5 nm) polymerizes into microtubules (right) (scale bar ~25 nm). (b) Highly ordered arrays of tryptophan amino acids that absorb radiation in the ultraviolet spectrum. Reproduced with permission from Ref.~\citenum{Celardo2019}. Copyright 2019 by IOP Publishing Ltd.}
  \label{FigBio1}
\end{figure}

\begin{figure}[ht!]
  \includegraphics[width=5.5in]{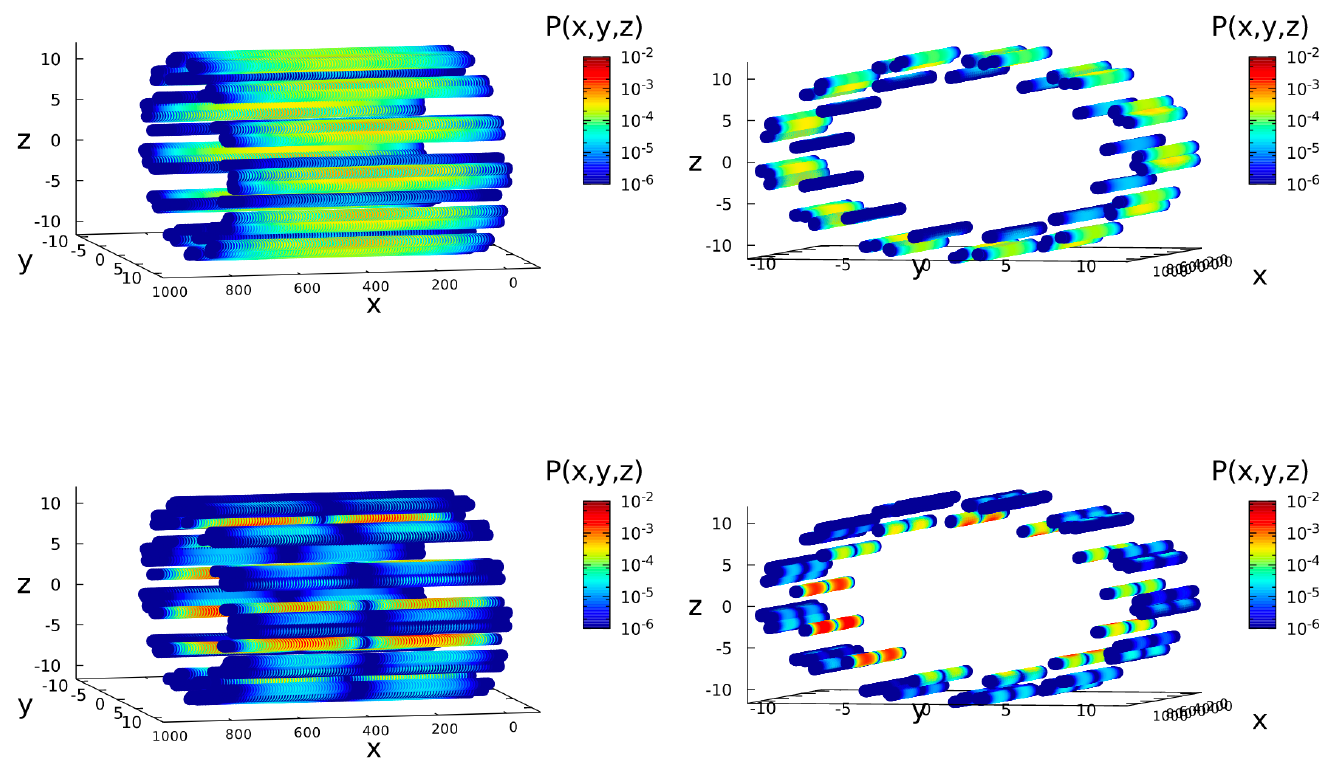} 
  \caption{The quantum probability of finding the exciton on a tryptophan of a microtubule segment of 100 spirals with 10,400 tryptophan molecules is shown for the extended superradiant state (top) in lateral view and (bottom) in cross section. Lengths are expressed in nm. Reproduced from with permission from Ref.~\citenum{Celardo2019}. Copyright 2019 by IOP Publishing Ltd.}
  \label{FigBio2}
\end{figure}

 Sensitive relationships between the chirality of underlying quantum (charge, spin, exciton, and plasmon) states and biological function abound. Kurian and colleagues have shown that certain chiral enzyme complexes with palindromic symmetry conserve parity,\cite{Kurian2016} and that the chirality is essential for the global synchronization of plasmon-like van der Waals fluctuations and for the symmetric recruitment of energy from DNA substrates for the site-specific formation of double-strand breaks. The application of tools from quantum optics to describe biological chromophore lattices has resulted in the recent prediction of ultraviolet SR in certain cytoskeletal filaments \cite{Kurian2016}, which exhibit a striking spiral-cylindrical chiral symmetry (\textbf{Fig.\ \ref{FigBio1}}) that is reflected in the excitonic wave functions distributed over the chromophore network (\textbf{Fig.\ \ref{FigBio2}}). The relationship between this electronic SR and its spintronic counterpart is an active area of investigation, which may lead to advanced biosensors and diagnostics. 

\par Exploiting the CISS effect to predict, control, and enhance  biological responses is a tantalizing possibility. Take, for example, the case of human immunity. 
T-cells initiate the body's immune response by interacting, \emph{via} their T-cell receptors, with major histocompatibility complex (MHC) peptides on antigen-presenting cells that were exposed to pathogens. MHC molecules are membrane-bound glycoproteins that form unusually stable bound configurations with antigenic peptide ligands (pMHC), displaying them on the cell surface  for recognition by T-cells \emph{via} T-cell receptor (TCR) engagement \cite{Rossjohn2015}. TCR activation promotes several signaling cascades that ultimately determine cell fate by regulating cytokine production, as well as cell survival, proliferation, and differentiation \cite{Courtney2018}. Studies have found that spin-polarized states of antigenic peptides may affect the ability of a TCR to recognize different peptides through conformationally-induced spin moments, rather than sheer topology-based affinity, a condition that would render the immune recognition process as fundamentally spin-specific \cite{Antipas2015,Antipas2015a}. 

Studies of Antipas \textit{et al.} show that different pMHCs with near-identical stereochemistries were complexed with the same TCR, resulting in distinctly different quantum chemical behaviors that depended on the peptide's electron spin density and expressed by the protonation state of the peptide's N-terminus
.\cite{Antipas2015} Spin polarization of different peptides can thus be correlated with downstream signal transduction pathways and their activations of biosynthesis at the transcriptional level.  Other studies have shown that noncovalent and dispersive interactions between biological molecules are critical to their functions, in which the electronic charge redistribution in chiral molecules is accompanied by spin polarization \cite{Kumar2017, Stoehr2019}. Studying chirality, spin polarization, and downstream response in pMHC-TCR interactions could improve our understanding of the impact of chiral-quantum effects on the human immune response, and to develop better tools for therapeutic intervention.

\section*{Chiral Degrees of Freedom in the Interaction of Matter and Electromagnetic Fields} \label{fields}

\subsection*{Chirality Imprinting}

\indent \par Handedness can be transferred from a chiral object to an achiral object or achiral medium, \textit{i.e.}, chiral molecules adsorbed on surfaces induce chirality on the substrate.  Indeed, we have found in our theoretical analyses that CD and optical rotatory dispersion signatures can be induced through chiral imprinting \cite{Goldsmith2006, Mukhopadhyay2007, Mukhopadhyay2009}. In fact, it is possible for the magnitude of the imprinted chiro-optical signature to be larger than that of the chiral dye itself.

\par A kind of handedness can be imparted to excited states using circularly polarized excitation. For example, molecules with degenerate x- and y- polarized excited states, when excited by left- \textit{vs.} right-handed circularly polarized light wave functions are complex conjugates of one another. If the excited state induces chemistry on a time scale that is fast compared to the time scale of decoherence, the chemistry may be imprinted by the handedness of the light excitation. For example, Skourtis \textit{et al.} predicted that electron transfer yields can be influenced by the handedness of circular excitation.\cite{Skourtis2008}  Combining circularly polarized excitation with chiral bridges can thus affect electron transfer chemistry.\cite{Bloom2018, Bloom2017}

\subsection*{Exploring Photonic Chirality through Orbital Angular Momentum}

\indent \par Photon packets can be described by several quantum numbers that describe their intrinsic spin angular momentum (SpAM), their number state, their energy, and their transverse spatial mode. For spin and transverse modes, the light can carry angular momentum which can interact both classically and quantum mechanically with materials. While SpAM is widely used for quantum information protocols \cite{Lloyd1999}, helical transverse chirality (orbital angular momentum, or OAM) is not commonly used to transmit nor to encode additional information in the emitted photon stream from a material \cite{Flamini2019}.

\par Similar to SpAM, OAM modes carry quantized angular momentum in proportion to the electric field phase difference around a centrosymmetric position. OAM can thus carry quantum information through pure and entangled states \cite{Fickler2012}. One exciting potential application of OAM will be the ability to drive transitions beyond the dipole limit \cite{Forbes2019}.  Many qubit transitions are defined within a quadrupolar or magnetic dipole subspace, and such transitions can only be driven by electric field gradients, carried in the OAM field. Improving the fidelity of OAM quantum information transfer may require near-field OAM photonics, in the visible and microwave regions \cite{Pu2015, Gorodetski2013}.   

\par Recently, researchers have found that OAM can induce specific quantized transitions in atomic ions \cite{Schmiegelow2016}, highly bound Rydberg excitons in semiconductors \cite{Konzelmann2019}, and potentially also in non-degenerate valley states of 2D materials \cite{Qiu2015}. These  experiments demonstrate the potential for helical OAM to drive states amenable for use in quantum information systems, and given the latter two cases, extended chiral excitons may be the key to unlocking the potential of chiral transverse modes in information transduction. 

The effect of OAM dichroism in isotropic matter has been theoretically predicted by Afanasev \textit{et al.} \cite{afanasev2017circular}. With this additional degree of freedom, the definition of dichroism becomes somewhat ambiguous. For instance, Kerber and collaborators in their paper on OAM-light interacting with nano-antennas \cite{kerber2018orbital} point out, that even for the case of fixed, yet non-zero OAM, one gets six distinctive types of dichroic effects. As a convenient solution to this branching problem, they presented the following probe classification: (1) parallel class beams with $|\textrm{OAM, SpAM}\rangle = |\uparrow \uparrow\rangle$ and (2) anti-parallel class beams $|\textrm{OAM, SpAM}\rangle = |\uparrow \downarrow\rangle$. The importance of this classification was earlier recognized in the realm of OAM light interacting with nano-objects \cite{zambrana2016far} and atomic matter \cite{quinteiro2017twisted}. This framework can be straightforwardly adopted in other subfields of twisted light and matter studies, for example electron, proton and neutron beams used as matter probes. 

Violation of spin selection rules due to the photon-atom OAM transfer was observed experimentally in photo-excited trapped ions. Later, based on these discoveries, the effect of local CD has been predicted on the level of multipolar contributions into an atomic photo-absorption amplitude \cite{solyanik2019excitation}. It has been concluded that OAM light of different helicities does not couple symmetrically to the atoms with high-order multipolar content, reflecting on the earlier findings. The phenomenon is impact-parameter-dependent and becomes stronger towards the beam axis. 


\section*{Chirality in the Quantum Sciences}\label{quantum}

\indent \par A scientific revolution enabled by quantum information processing is underway. Successful quantum computing hardware will depend both on incremental technological improvements and on disruptive applications of physical laws underpinning how qubits read, store, and transduce information. The challenge addressed here is to rethink quantum information protocols to incorporate chiral `handles', that is: Can nanoscale chirality be leveraged to create innovative approaches to quantum information processing (QIP)?
\par Exploiting chirality as a design framework for quantum devices will allow the control of spin, charge, and energy transport using molecules and interfaces so that quantum information can be preserved and transferred at room temperature. A key concept is the manipulation of the magnetic response in chiral molecules and engineered nanomaterials \emph{via} CISS. This approach includes controlling spin filtering and polarization capabilities of molecules and engineered nanomaterials, and generating qubits through molecular design, surface architectures, and tailored interactions with light with chiral degrees of freedom. A complementary focus is the study of quantum transduction processes at soft--hard material interfaces, involving the transfer of both spin polarization from electrons to nuclei and field-mediated chirality transfer. 


\subsection*{Spin Superradiance and Chiral-Induced Spin Selectivity}

\indent \par Cooperative effects arise from the collective behavior of the constituents of a system, and therefore they are associated with the system as a whole and not with its individual components. These phenomena occur at every scale, ranging from the structure of atoms in crystals to ferromagnetism, superradiance (SR), superconductivity, functionality in complex molecules, and the emergence of life from biomolecules.\cite{Fleming2008} One of the most interesting properties of cooperative effects is their robustness to noise induced by external environments. For this reason, cooperative effects could play essential roles in the successful development of scalable quantum devices that operate at room temperature. 

\par A well known example of a robust cooperative effect is superconductivity, but other quantum cooperative effects, such as SR, were also shown to be robust against noise \cite{Giusteri2015, Celardo2014}. Superradiance,  proposed by Dicke in 1954 \cite{Dicke1954}, arises from the excitation of an ensemble of individual two-level systems and results in an emissive, macroscopic quantum state. Superradiant emission is characterized by an accelerated radiative decay time, where the exponential decay time of spontaneous emission from the uncoupled two-level system is shortened by the number of coupled emitters. In addition, when the excitation is incoherent \cite{Bonifacio1975}, SR exhibits a delay or build-up time during which the emitters couple and phase-synchronize to each other. This time corresponds to the time delay between the excitation and onset of the cooperative emission. In the case of SR, the coupling of an ensemble of emitters to an external field can induce an energy gap, making superradiant states robust to disorder. Interestingly, the superradiant energy gap, in certain limiting cases, is the same as the superconducting gap \cite{Chavez2019}. 

\par SR has been observed in a variety of systems \cite{Cong2016}, including cold atomic clouds \cite{Araujo2016}, photosynthetic antenna complexes \cite{Monshouwer1997}, molecular aggregates,\cite{DeBoer1990, Fidder1990} QDs,\cite{Scheibner2007, Brandes2005} nitrogen vacancies in nanodiamonds,\cite{Bradac2017} and lead--halide perovskite nanocrystal superlattices.\cite{Raino2018} 
This effect is relevant to enhance absorption and  energy transfer, which was proposed to improve the efficiency of light-harvesting systems and photon sensors.\cite{Higgins2014,Hu1997,Struempfer2012,Hu1998,Sener2007} SR also leads to spectrally ultra-narrow laser beams.\cite{Bohnet2012}

\par Although the vast majority of systems exhibiting SR involve electronic transitions at optical frequencies, SR has also been observed in spin systems.\cite{Kiselev1988} Spin SR has attracted much attention due to its many possible applications to sensing and spin masers (microwave amplification by stimulated emission of radiation).\cite{Yukalov2007,Angerer2018,Rose2017,Li2018x} Measuring the delay time of a superradiant burst provides an accurate evaluation of the triggering intensity because the delay time is exponentially dependent on the intensity of the external pulse and spin SR can be triggered by extremely weak external pulses. Thus, systems exhibiting spin SR can be used to produce sensitive detectors. Another very interesting application of spin SR is spin masers.\cite{Jin2015,Breeze2018,Salvadori2017} A superradiant spin system is a source of coherent radiation at radio frequencies between 0.3 and 300 GHz and a wavelength between 1 m and 1 mm, and can act as the microwave analogue of the laser. There are important applications for masers in ultrasensitive magnetic resonance spectroscopy, astronomical observation, space communication, radar, and high-precision clocks.

\par The key to spin SR is the population inversion of the emitters. In spin systems, population inversion corresponds to spin polarization, which is a required condition for SR emission and maser operation when a polarized spin population is present in a  microwave resonator cavity. For this reason, it would be intriguing to exploit the CISS effect \cite{Zoellner2020, Zoellner2020b} in connection with spin SR and spin masers. Indeed, the spin-polarized beam emerging from chiral molecules due to the CISS effect could be used to induce an SR pulse. A beam of polarized electron spins could also be used to operate a spin maser when coupled to  a  microwave resonator cavity. Moreover, it is known that a polarized electron beam couples with nuclear spins by hyperfine spin-spin  interactions. This coupling would produce a shift of the nuclear magnetic resonance frequency of nuclear spins and would thus enhance the coupling between the nuclear spins and the resonator. These effects could be used both to study the CISS effect and to build more efficient SR or maser nuclear spin systems. 

\subsection*{Controlling Spin Polarization and Entanglement in a Hybrid Chiral Molecule/Quantum Dot System}

\indent \par
A recent study found that the CISS effect vanishes when all electron states with the same energy are equally likely, a consequence of the Onsager reciprocal principle.\cite{Dalum2019} The generality of this result means that the CISS effect needs to be understood in terms of the specific experimental settings. Three possible situations have been suggested \cite{Dalum2019}: the electronic  states with the same energy not being equally probable (\textit{e.g.}, for electrons generated optically by a laser, due to selection rules inherent in photoexcitation processes); the presence of accidental degeneracy in the molecular spectrum, which enhances the SO coupling; or the presence of a magnetic lead. More recently, an analysis based on symmetry in electronic transmission was carried out to better understand the origin of the  CISS effect~\cite{Zoellner2020}.  

\begin{figure}[ht!]
\centering
\includegraphics[width=3.5in]{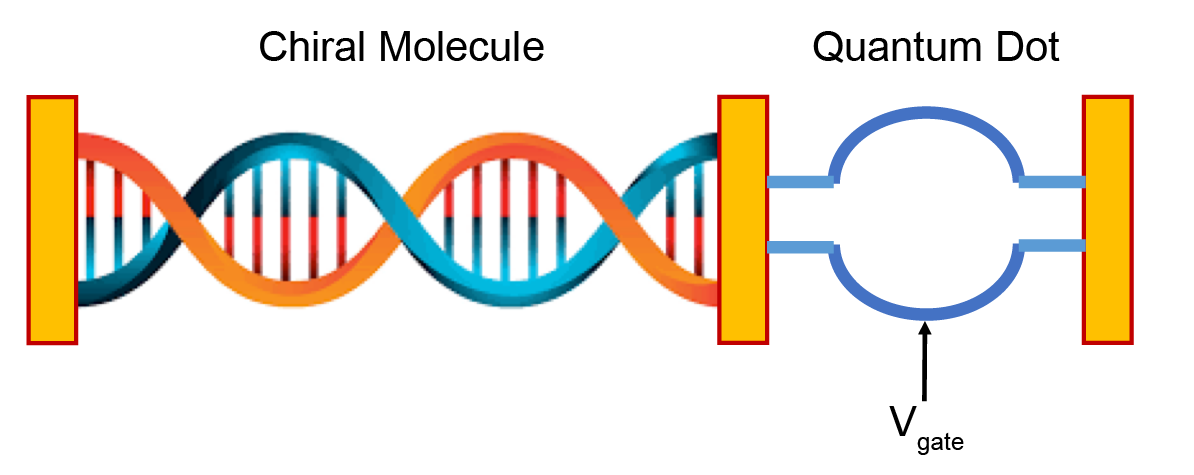}
\caption{A schematic illustration of a `super chiral molecule' -- a coupled chiral-molecule/quantum-dot hybrid structure, for controlling and manipulating spin polarization and entanglement. The gate voltage is applied to change the QD structure for controlling spin polarization.}
\label{FigCISS_hybrid}
\end{figure}

\par While strengthening the SO coupling by coupling a chiral molecule to a heavy metal or to a superconductor can enhance spin polarization, control and manipulation are difficult. A potential approach is to exploit a hybrid structure that couples a chiral molecule to a two-dimensional (2D) QD to form a `super chiral molecule' to control and manipulate spin polarization, as illustrated schematically in \textbf{Fig.\ \ref{FigCISS_hybrid}}. The idea originates from recent studies to control spin polarization by exploiting classical chaotic dynamics \cite{Ying2016a,Liu2018} and spin Fano resonances \cite{Liu2020}. 

\par The role of classical dynamics in spin transport is intriguing from the point of view of a classical--quantum correspondence, as spin is a purely relativistic quantum mechanical variable without a classical counterpart. Nevertheless, due to SO coupling and because the orbital motion does have a classical analogue, the nature of classical dynamics can influence spin. Full quantum calculations and a semiclassical theory revealed that spin polarization can be modulated effectively if the geometrical shape of a QD can be modified to produce characteristically distinct classical behaviors ranging from integrable dynamics to chaos.\cite{Liu2020} Chaos can play distinctive roles in affecting spin polarization, depending on the relative strength of the SO coupling. For weak couplings with characteristic interaction lengths much larger than the system size, chaos can preserve and even enhance spin polarization. In the strong coupling regime, where the interaction length is smaller than the system dimensions, chaos typically degrades or even destroys spin polarization. In 2D materials, such as graphene, a QD can be realized by applying a properly designed gate potential. The total spin polarization from the hybrid structure can then be manipulated electrically.

\par In electron transport through mesoscopic systems, the various resonances associated with physical quantities -- such as conductance and scattering cross sections -- are characterized by the universal Fano formula \cite{Huang2015}. Recently, a Fano formula was discovered to characterize the resonances associated with two fundamental quantities underlying spin transport: the spin-resolved transmission and the spin polarization vector.\cite{Liu2020} The Green's function formalism was generalized to describe spin transport and the Fisher-Lee relation was used to compute the spin-resolved transmission matrix, enabling the spin polarization vector to be calculated and leading to a universal Fano formula for spin resonances. The resonance width depends on the nature of the classical dynamics, as defined by the geometric shape of the dot, and this property could be exploited for control. Since characteristically distinct types of classical dynamics including chaos can be readily generated in the QD through geometric deformations \cite{Huang2018}, Fano spin resonances can be modulated accordingly. This work provides a theoretical foundation for the general principle of controlling spin polarization in the chiral molecular system through manipulating the classical dynamics in the QD, which can be experimentally realized by applying a properly designed local gate potential profile. Likewise, modulating the classical dynamics in a different way can enhance the spin Fano resonance. The control principle was computationally demonstrated in a key component in spintronics: a class of nanoelectronic switches, where the spin orientation of the electrons associated with the output current can be controlled through weakening or enhancing a Fano resonance.\cite{Liu2020} 

\subsection*{Quantum Information Storage and Transduction}

\indent \par We believe it is timely to exploit the distinctive properties of chiral molecules and chiral material interfaces to develop room-temperature device technologies in quantum sensing, quantum storage, and quantum computing. Tailoring OAM couplings by acting either on the spin quantum states (that serve as qubits) and/or on the interaction potentials could enable room-temperature transduction of quantum information. 
\par Quantum information processing and quantum sensing have made enormous strides in the past decades. Superconducting circuits and ion traps have led to a race for supremacy in quantum computing,
while ultracold atoms and molecules, nitrogen-vacancy and other color centers have enabled a variety of approaches to quantum sensing (\textit{e.g.}, magnetometry, dark-matter detection, atomic clocks).
With the exceptions of NV centers and atomic vapors, experiments require extremely low temperatures (mK range) to protect the coherent states and to allow sufficient time for entanglement to evolve, or for measurements to be made. This constraint restricts their use to systems with sizable infrastructure support. The main drawbacks of qubits based on defect centers at present are the difficulties of fabricating them at specific locations, selectively addressing different centers, and establishing quantum information transfer beyond the short dipole--dipole interaction length scales.

\par In contrast to the approaches described above, chemical synthesis of chiral materials affords the opportunity to build quantum information systems from the bottom up, taking full advantage of the quantum properties of matter on the atomic scale. Chiral molecular systems differ from current qubit implementations described above in three important ways. First, chemical synthesis enables control over the nature of the qubit itself, thus enabling careful tuning of individual quantum states. Second, covalent and non-covalent interactions between molecules can be used to construct atomically precise arrays of qubits. This approach offers the possibility of controlling and interrogating the properties of a qubit both in isolation and as part of an array, providing insights into the quantum properties of multi-qubit arrays. Third, chiral molecules have the potential to serve as long-range quantum information transducers.

\par It may be possible to transfer chiral spin information between electrons and nuclei \textit{via} hyperfine interactions.
Spin-polarized electrons can be generated on the surfaces of topological insulators (TIs) through the application of an electrical current \cite{Tian2017}. The coupling of the electron spin to the nuclear spin \emph{via} hyperfine interactions highlights the ability to produce dynamic nuclear polarization \cite{Tian2017, Sharma2017}. A promising application is a rechargeable spin battery \cite{Tian2017}. 
These recent results point to two applications: \textbf{(1)} TIs can be used to generate spin-polarized electrons, which can interact with nearby chiral molecules and biomolecular structures through highly reproducible, high-precision contacts; \textbf{ (2)} Interactions with radical pair states are also possible, particularly in the context of the repair of lesions in duplex DNA, where the repair yield is dependent on the strength and angle of the applied magnetic field \cite{Zwang2018}. Chiral spin modes on the surface of a TI can be used as an additional degree of freedom for manipulating the polarization.\cite{Kung2017}


\par
\begin{figure}[ht!]
  \includegraphics[width=5in]{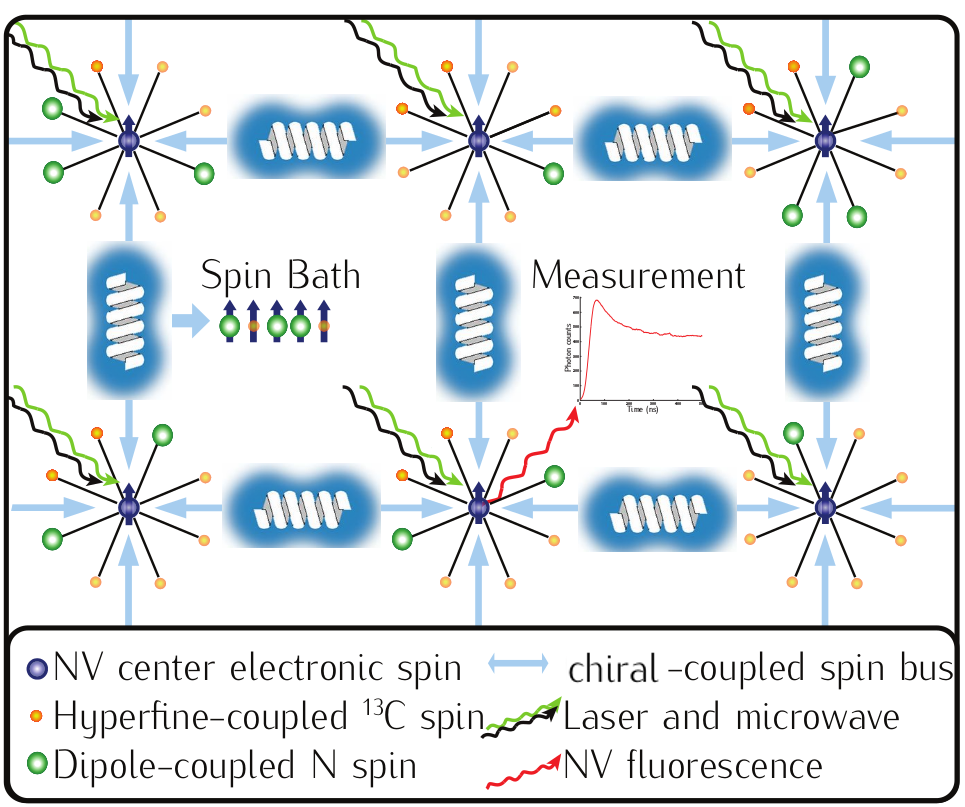} 
  \caption{\textbf{Chiral molecules could be used as interconnects or quantum information transducers that have a longer-range than dipolar-coupled spin buses}.} 
  \label{FigChiralConnects}
\end{figure}

 Chiral molecules could also be harnessed for quantum information transduction.  Nanochiral materials could be used as quantum wires connecting a node of quantum sensors. For example, we envision testing chiral materials as tractable, in-chip solid-state quantum wires connecting established quantum sensors (\textit{e.g.}, color centers in diamond, silicon, or silicon carbide), see \textbf{Fig.\ \ref{FigChiralConnects}}. Proposals for connectivity among quantum sensors has traditionally relied on dipole-coupled spin buses,\cite{Doherty2016} which unfavorably limit the maximal distance between nodes, and which can hardly be engineered. Conversely, quantum information transduction through chiral materials will overcome both of these limitations, and have already been shown to be capable of quantum information transduction and topological-like transport over longer distances and in complex environments.\cite{Goehler2011}.

\subsection*{Decoherence and Entanglement Considerations}

\indent \par The realization of devices for near-room-temperature, CISS-based quantum information will necessitate a deep understanding of the decohering bath. Investigation of single-qubit coherence properties under controlled noise (electron--electron, electron--phonon scattering) environments will be needed to inform the design of chiral materials and devices for quantum information processing. For example, spin textures  are complex results of electronic correlations.\cite{Tao2018} If electron--electron interactions can be effectively decoupled, CISS can still be sustained by the material, since it results essentially from the SO coupling and breaking inversion symmetry under electron transfer, transport, or polarization conditions. Understanding the bath dynamics and the application of reservoir engineering principles can be achieved with proof-of-principle experiments relying on controlled charge injection into nanochiral materials, combined with quantum information/sensing protocols for spin-state preparation, control, and readout.

\par Another key issue is entanglement. For a quantum system with multiple degrees of freedom, loss of coherence in a certain subspace is intimately related to the enhancement of entanglement between this subspace and another one. It would be useful to investigate intra-particle entanglement between spin and orbital degrees of freedom in the chiral molecular system with different types of classical dynamics in the QD. Of particular interest is the spin degree of freedom in the weak SO coupling regime where, as existing studies suggested, it is possible to use classical chaos to  enhance SO entanglement significantly at the expense of spin coherence. \cite{Ying2016a,Liu2018,Liu2020}

\subsection*{Control of Light--Matter Interactions}

\par Taken together, orbital and spin angular momenta provide exciting avenues for quantum control. The critical challenge for using OAM for quantum control is to enforce the interaction of the emitter with the inhomogeneous spatial mode of the electromagnetic field, which is addressed by the following two questions: 

\textbf{(i)} How can one design materials that interact strongly with orbital angular momentum fields \textit{via} extended excitonic states? Many delocalized systems already couple strongly to spin angular momentum through either macroscopic helicity (\textit{e.g.}, helical molecular aggregates) or through degenerate points in their band structure (\textit{e.g.}, valleytronic 2D materials), but spin angular momentum combined with orbital angular momentum remains experimentally underexplored. It needs to be established how mesoscopic helicity in designed and self-assembled materials can control and enhance interactions with orbital
angular momentum fields. 

\textbf{(ii)} How can one implement sub-wavelength optical fields with orbital angular momentum that will strongly interact with matter? Near-field photonics can create field gradients that are far stronger than the free-space modes (thus preventing decoherence) and that can be used to drive high-order transitions (such as electric quadrupole) far more efficiently than the vacuum (\textit{i.e.}, spontaneous emission).  Controlling quantum information \textit{via} spatially engineered electromagnetic fields that provide environmental control of angular momentum would be transformative for quantum-enabled technologies.

Let us consider light-matter interactions akin to deformations in chiral molecules induced by mechanical means such as applying an external force. These deformations are
 nontrivial due to the non-uniform distribution of the load depending on the molecule structure. A model by de-Gennes \cite{deGennes2001} predicted that DNA externally pulled by mechanical means deforms only close to the mechanical contacts, leaving the bulk of the molecule undeformed. Such a stretching process is thus hampered by details of the contact. An alternative to pulling on the molecule is to shine light upon it at low intensity and at an appropriate frequency to modulate the strength of the electronic bonding in the molecule and thus changing its rigidity at a fixed external force so that it stretches or compresses in particular ways in the bulk structure. This approach is also known as stretch engineering. In addition, the role of symmetries in the CISS effect has recently been considered in Ref.\ \citenum{Sierra2019} where the authors present a model which consists of two inter-connected tight-binding chains, mimicking two interacting helices, including spin–orbit interaction and attached to two fermionic reservoirs playing the role of current terminals. It is remarkable that SO coupling is particularly sensitive to such mechanical changes being dependent on the detailed relative geometry of the spin-active units \cite{HuertasHernando2006, Medina2016}. Indeed, deformation dependent spin activity has been shown experimentally \cite{Kiran2017, Vetter2020} and has been modeled analytically in DNA \cite{Salazar2018}  and in oligopeptides \cite{Torres2020} where the role of hydrogen bonding as the model is stretched or compressed has been addressed.
 
A natural extension of the previously considered works is to model the non-equilibrium spin response induced by the interplay of both SO interactions and light-matter coupling in these chiral systems. The natural formalism for dealing with periodically driven interactions is the so called Floquet theory \cite{Oka2019}, analogous to the Bloch theory for spatially periodic interactions. Within this context, considering the coupling of charge carriers in the chiral sample to monochromatic radiation fields could provide another means to address the spin degree of freedom, offering additional tunability to the CISS effect. The use of the Floquet formalism in the description of laser-assisted transport in molecular junctions has been discussed \cite{Urdaneta2007}.

\begin{figure}[!ht]
  \includegraphics[width=6.5in]{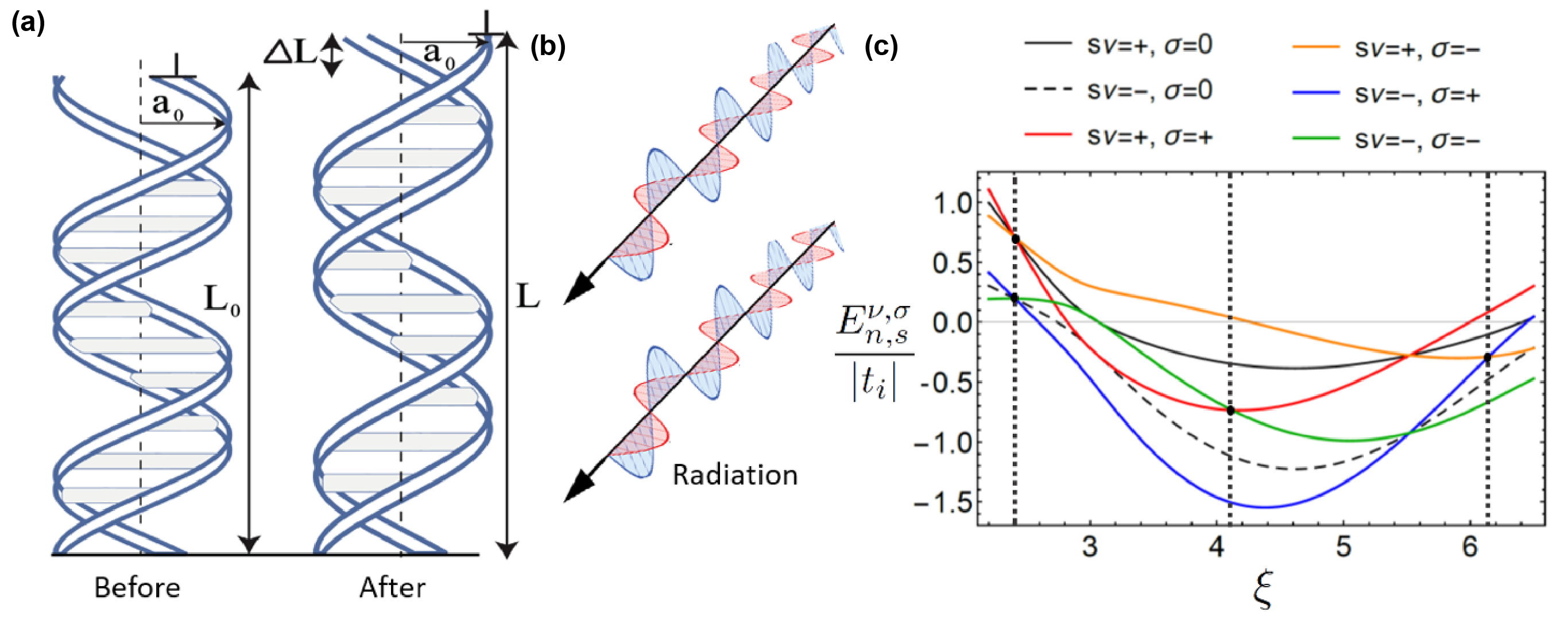}
  \caption{Effects of light on chiral molecules through the Floquet approach to achieve stretch and spectral engineering a) Molecules can be made to stretch or contract by impinging light modulating the SO coupling (modified figure from Ref. \citenum{Salazar2018}). b) Spectrum normalized by base pair to base pair overlap strength as a function of spin state (s), transport direction ($\nu$) and pseudo-spin quantum ($\sigma$) \textit{vs.}\ the normalized frequency ($\xi$) of impinging light.}
  \label{FigMolLight}
\end{figure}

From a physical point of view, linearly polarized electromagnetic (EM) radiation carries no angular momentum, in contrast to the circularly polarized EM fields. One can also consider chiral photons, \textit{i.e.}, photons carrying orbital angular momentum \cite{Ye2019}. Consequently, in the presence of linearly polarized radiation fields, the spin degeneracy of the charge carriers can be broken only if Rashba SO interaction is absent. In other words, circularly polarized electromagnetic fields break the spin degeneracy without SO interaction. For the case of linear dispersion, circular polarization is known to open gaps in the quasi energy spectrum at the degeneracy points when compared with linearly polarized EM fields \cite{Scholz2013}. Moreover, in dealing with the transport regime, one key point to be considered is that the distribution function of the Floquet states needs not to be a Fermi distribution function. The Floquet problem in which the non-equilibrium distribution is taken into account (already in the absence of static external magnetic field) is difficult and often one resorts to arguments that justify why the non-equilibrium distribution function can be replaced by a quasi-equilibrium (Fermi) distribution. Sometimes, the conditions that justify this replacement are not explicitly stated in the literature \cite{Peralta2018}. A possible means to model the photo-induced quasi one-dimensional spin transport in the DNA sample could be achieved by using the approach given in Ref. \citenum{dalLago2015}, where the driven Su–Schrieffer–Heeger model of polyacetylene is explored, and it is shown that competing effects among photon assisted processes and the native topology of the undriven system lead to nontrivial Floquet topological insulating phases \cite{Rudner2019}. Additionally, effective Hamiltonians within the so called off resonant regime allow for a physical description of the spin transport by means of exactly solvable models \cite{Lopez2015, Lopez2020}. Thus, within the Floquet formalism it becomes possible to deal with periodically driven interactions, and consider the effects of radiation on bonds using the formalism of stretch engineering. Other effects seen in low dimensional systems such as the spin ratchet effect could also be explored.

\section*{Future Outlook and Conclusions}

\indent \par In this perspective, we have reviewed the interplay between chirality and quantum effects in several contexts. We have discussed the role of diverse molecules, materials, and systems in which electromagnetic chiral degrees of freedom can be harnessed for spintronics and quantum information applications. When chirality is introduced into measurements, it produces quantifiable effects in electrochemical or redox reactions and in spin-selective conduction detected by scanning probe or break junction experiments. These effects are described by the CISS effect, which can be thought of as a spin polarization effect in the absence of magnetic fields. Theoretical descriptions of the CISS effect and other related observations have been done using combinations of DFT and tight-binding approaches, but it is still challenging to capture all the various effects involving electrons, spin, symmetry, and geometry especially for non-trivial systems of large numbers of atoms. Some of the most intriguing observations of SO interactions, the CISS effect, and other chiral-quantum phenomena have been in engineered chiral materials. These have included combinations of organic molecules and metal substrates, substrates with heavy elements, 2D materials functionalized with chiral molecules, and hybrid structures of chiral molecules and nanoparticles. Related effects have also been observed in biological structures. Chirality can also be imparted to electromagnetic fields and can be observed in OAM modes, which may have applications in quantum information systems. Throughout all these contexts, the interplay between chirality, quantum effects, and materials grants new possibilities for controlling spin, charge, and energy transport for quantum information processing.
Moving forward, chiral-enhanced nanoscience will require a multidisciplinary effort combining cutting-edge materials design and characterization with diverse theoretical strategies, with the long-term goal of designing and controlling chiral (qubit) spin states that operate at room temperature. Some of the crucial developments that will be needed in the field to further advance the understanding and application of chirality and quantum effects in quantum information processing, storage and transduction include:

\begin{itemize}
\item Exploration of novel materials such as topological chiral materials, 2D materials and heterostructures, 2D magnets, etc.;
\item Design and evaluation of artificial chiral materials that can be used to control electronic, magnetic, and optical effects;
\item Integration of multiple categories of materials such as organic molecules and conjugated systems, 2D quantum materials, chiral surfaces, thin films, and nanomaterials;
\item Fundamental probing of biological processes to understand mechanisms that can be brought into artificial systems;
\item Improved modeling of chiral systems that go beyond DFT, particularly for  larger molecular and hybrid systems that more closely resemble experimental conditions;
\item Design, implementation, and testing of new quantum device architectures to take advantage of chiral-quantum effects;
\item Matching chiral materials that exhibit quantum effects with properties suitable for existing quantum device architectures.
\end{itemize}

Based on the exciting work that we have reviewed here, and the extensive ongoing efforts in many groups, we anticipate that materials and systems that take advantage of inherent chiralities and interactions with other quantum effects will play an important role in the next revolution in quantum devices.

\begin{acknowledgement}
E.J.G.S. acknowledges the EPSRC Early Career Fellowship (EP/T021578/1) and 
the University of Edinburgh for funding support. C.H. acknowledges funding by Deutsche Forschungsgemeinschaft (DFG) \emph{via} the project ``Structure--property relationships for SO effects in chiral molecules'' (HE 5675/4-1). A.D., R.G. and G.C. acknowledge financial support from the Volkswagen Stiftung (grant no. 88366) and from the the German Research Foundation within the project Theoretical Studies on Chirality-induced Spin Selectivity (CU 44/51-1). V.M., W.T.P., and Q.H.W. acknowledge support from the National Science Foundation Quantum Leap Challenge Institutes (QLCI-CG-1936882).

\end{acknowledgement}

\bibliography{ACSNano.bib}
\end{document}